\documentclass[useAMS,usenatbib,usegraphicx]{mn2e}

\def\aap{A\&A}
\def\apj{ApJ}
\def\apjs{ApJS}
\def\apjl{ApJL}
\def\mnras{MNRAS}

\def\aaps{A\&A Supp.}
\def\araa{ARA\&A}       
\def\apss{ApSS}
\def\jcap{JCAP}

\usepackage{amsmath}
\usepackage{natbib}
\usepackage{epsfig}
\usepackage{txfonts}
\usepackage{url}
\usepackage{graphicx}
\usepackage{dcolumn}
\usepackage{amsfonts}
\usepackage{amssymb}
\usepackage{bm}
\usepackage{hyperref}
\usepackage{hypernat}
\usepackage{url}
\usepackage{subfigure}
\usepackage{txfonts}
\usepackage{color}
\usepackage{enumitem}

\usepackage{float}

\title[LIL CMB component separation]{Linearized iterative least-squares (LIL): 
A  parameter fitting algorithm for component separation in multifrequency CMB
experiments such as Planck}
\author[Khatri]{Rishi Khatri$^{1}$\thanks{E-mail:
khatri@mpa-garching.mpg.de}\\
$^{1}$ Max-Planck Institut f\"ur Astrophysik, Karl-Schwarzschild-Str. 1,
D-85740 Garching, Germany
}
\begin{document}

\newcommand{\Acmb}{{{A_{\rm CMB}}}}
\newcommand{\Alf}{{{A_{\rm lf}}}}
\newcommand{\dellf}{{{\delta_{\rm lf}}}}
\newcommand{\delblf}{{{\delta^{\rm lf}_{\beta}}}}
\newcommand{\flf}{{{f_{\nu}^{\rm lf}}}}
\newcommand{\nulf}{{{\nu_{0}^{\rm lf}}}}
\newcommand{\blf}{{{\beta_{\rm lf}}}}
\newcommand{\Adst}{{{A_{\rm d}}}}
\newcommand{\deldst}{{{\delta_{\rm d}}}}
\newcommand{\delbdst}{{{\delta^{\rm d}_{\beta}}}}
\newcommand{\fdst}{{{f_{\nu}^{\rm d}}}}
\newcommand{\nudst}{{{\nu_{0}^{\rm d}}}}
\newcommand{\bdst}{{{\beta_{\rm d}}}}
\newcommand{\Tdst}{{{T_{\rm d}}}}
\newcommand{\Aco}{{{A_{\rm co}}}}
\newcommand{\fco}{{{f_{\nu}^{\rm co}}}}
\newcommand{\kB}{{{k_{\rm B}}}}
\newcommand{\changeR}[1]{\textcolor{red}{#1}}

\maketitle

\begin{abstract}
We present an efficient algorithm for the least squares parameter fitting
optimized for
component separation in multi-frequency CMB experiments. We
sidestep some of the problems associated with non-linear optimization by
taking advantage of the quasi-linear nature of the foreground model. We demonstrate our algorithm, linearized
iterative least-squares ({\sc LIL}), on the publicly available Planck sky model
FFP6   simulations and compare our result with the other algorithms. We
work at full Planck resolution and show that degrading the resolution of
all channels to that of the lowest frequency channel is not necessary.
 Finally we present results for the publicly available Planck data.
Our algorithm is extremely fast, fitting 6 parameters to 7 lowest Planck
channels at full resolution (50 million pixels) in less than
160 CPU-minutes (or few minutes running in parallel on few tens of
cores). {\sc LIL} is therefore easily scalable to future experiments which may
have even higher resolution and more frequency channels. We also naturally propagate the
uncertainties in different parameters due to noise in the maps as well as degeneracies
between the parameters to the final errors on the parameters using Fisher
matrix. One indirect application of {\sc LIL}
could be a front-end for Bayesian parameter fitting to find the maximum of
the likelihood to be used as the starting point for the Gibbs sampling. We
show for rare components, {such as the carbon-monoxide emission,} present in small fraction of sky, the optimal approach should
combine parameter fitting with model selection. LIL may also be useful in
other astrophysical applications which satisfy the quasi-linearity criteria.
\end{abstract}


\section{Introduction}
Multi-frequency CMB experiments such as WMAP \citep{wmap} and Planck \citep{planck} make it possible to
separate the observed signal into CMB and foreground components by taking
advantage of the fact that different components have different spectral
properties. Component separation allows us to use a larger fraction of the
sky for cosmological analysis. For high sensitivity experiments, such as
Planck, even in the relatively clean parts of the sky the foregrounds would be
above the noise level and we need component separation to take advantage of
the full sensitivity of the experiment. Although most component separation
methods assume quite simple foreground models, the foregrounds in reality
are much more complicated. We are limited to simple models mostly because
of the lack of spectral resolution in the CMB experiments. Planck has just
7 frequency bands from 30 GHz to 353 GHz where CMB dominates over a
significant portion of the sky  and an additional two bands at 545
GHz and 857 GHz which are dominated by the foregrounds over most of the
sky. Foregrounds on the other hand are expected to have multiple
components such as synchrotron \citep{haslam}, free-free \citep{wmapfree}, anomalous dust
emission (\citealp{erickson}; \citealp{dl1998,dl1998b,dl1999}),
{carbon-monoxide (CO) line emission \citep{dame2001,magnani1985,hartmann1998,magnani2000}},
dust emission (\citealp{iras}; \citealp{cobedmr}; \citealp{cobedust};
\citealp{cobefiras}; \citealp{sfd1998}; \citealp{sfd1999}).   Each of these components is in fact made up of
superposition of emission from different regions along the line of sight
with different temperatures, spectral indices and intensity. In addition
there is also a contribution from the extra-galactic radio
(\citealp{ls1972}; \citealp{gervasi2008}) and infrared
backgrounds \citep{hauser1998}. {The
cosmologically interesting components present are the CMB and the thermal
Sunyaev-Zeldovich (SZ) effect \cite{zs1969,sz1972}. In this paper we will only
consider the CMB component and apply our method to the SZ effect separation
in a separate publication \citep{k2015,ks2015}.}

The  component separation methods  must therefore make many simplifying
assumptions about the foregrounds. Although the details of different
methods differ, broadly they can be classified according to whether they
assume that the emission laws vary over the sky or not. For example, the template fitting method
 Spectral Estimation Via Expectation
Maximization ({\sc SEVEM}) (\citealp{sevem}; \citealp{sevem2}) makes the assumption that the spectral parameters are
constant over large regions of the sky so that a template subtraction can
be performed. The Spectral Matching Independent Component
Analysis ({\sc SMICA}) \citep{smica} and its Bayesian incarnation \citep{bayessmica} construct and fit a template or model of foreground emission, CMB
and noise to the channel maps in the spherical harmonic domain and
also makes the assumption that the emission laws do not vary with
 the spherical harmonic mode numbers $\ell,m$ i.e. they are constant over the whole sky.
{\sc Commander} \cite{eriksen2006,eriksen2008} uses a simplified parametric model
motivated by the known physics about the foregrounds. The spectral
parameters vary from pixel to pixel on a much lower resolution map and
this method is therefore more flexible in modelling the varying emission
laws on the sky and a little closer to the physical
nature of the foregrounds. The Needlet Internal Linear Combination
algorithm ({\sc NILC}) \cite{nilc} constructs an internal linear combination
map in a wavelet space whose basis functions are needlets. The needlets are
localized both in the spatial and harmonic space and this method  lies
somewhere in between 
 the {\sc SEVEM} and {\sc SMICA} in terms of the basis. It fits the foregrounds as a function of both spatial
location as well as angular scale. This method also assumes 
that the emission laws over large portions of the sky are constant. 

The foregrounds provide useful information about the
galactic and extragalactic physics and accurate estimation of the
foregrounds components is therefore also important.  The  component
separation methods referred to above, except for {\sc
  Commander}, do not separate different foreground components but only the
sum of the foregrounds from the CMB. {In addition to these methods used by
Planck collaboration for the main CMB analysis as well as Sunyaev-Zeldovich
effect maps \cite{plancky} which also employs a modified ILC method \cite{milca}, several other  have been proposed over
the past two decades
\citep[e.g. ][]{bedini2005,maino2002,hobson1998,stolyarov2002,bobin2007,hansen2006,bonaldi2007}.
Most of these component separation methods  are reviewed in
\citet{leach2008}. We will not examine all the algorithms in detail but
just state the important difference from our algorithm that they all share the property of assuming the slowly varying
(with respect to the position on the sky)
emission laws for the foreground components.}

It is a reasonable question to ask if the assumptions about the foregrounds
and the CMB can bias our CMB results. {For example,}  can the foreground
cleaning/component separation introduce anomalies in the resulting CMB maps
or, what may be even worse, mask some of the primordial anomalies?
  It is therefore worthwhile to have different
foreground removal methods which operate on different assumptions. But more
importantly we need methods which rely on minimum amount assumptions about
the nature of foregrounds and the CMB, especially with regards to the
angular correlation structure of both the amplitudes and the emission laws, and only use what we physically know about the
foregrounds. {We will not attempt to answer all the questions raised in
  the previous paragraphs in this paper. We should however mention one
  particular application where the assumptions about the spatially slowly
  varying foregrounds fail but leave the examination of the consequences
  in detail to a separate paper \cite{k2015}.}

{The gas and
  dust in
  the galaxy, responsible for  the galactic foregrounds are concentrated
  into HI \cite{HB1997,hkb1996,dl1990} and molecular clouds\cite{dame2001} with angular dimensions on the
  sky much smaller than a degree. The molecular clouds have been detected
 away from the galactic plane at latitudes at high as $|b|=55^{\circ}$ \cite{magnani1985,hartmann1998,magnani2000}. Moreover the emission laws are functions
  of conditions in these clouds, such as temperature, which can vary even inside the
  clouds. Physically therefore we expect that the emission laws would vary
  from pixel to pixel at the resolution of Planck. It may well be that these
  variations are small enough that the assumptions in the currently
  employed component separations are valid. But they are still assumptions
  and need to be explicitly tested. Also the assumptions which may be
  reasonable for  the CMB anisotropy power spectrum may well fail for other
cosmological applications. One such application, where our method has
certain advantages, is looking for weak and
rare but concentrated signals in the Planck maps such as CO emission and
Sunyaev-Zeldovich effect in clusters. Our method provides a simple
canonical quantitative
measure of how good the assumptions, in our case the parametric model,
are in describing the data in the form of residuals between the data and
the model, or the $\chi^2$. We will use this additional information to
select between the different parametric models to check if CO emission is
present in a pixel or not. We will use the algorithms that we develop here
to separate out the CO contamination from the Sunyaev-Zeldovich effect
maps in a separate publication \cite{k2015}.}

The purpose of the present work is to present our algorithm and validate it. Our approach
is parameter fitting in pixel space, similar to the {\sc Commander}, but we
do so at full resolution relaxing the assumption  that
the spectral parameters are constant on degree scales. 
We
find that there is no  need to smooth the resolution of all maps to the lowest
resolution map. In fact re-beaming the low resolution maps to higher
resolution, as is done by {\sc SMICA} and {\sc NILC}, would be a better
choice yielding CMB maps with an effective  beam closer to that of the
highest resolution channels.  {\sc Commander} explores the
full posterior distribution using Gibbs sampling \cite{eriksen2006,eriksen2008} and is
therefore computationally very expensive, taking 100s per pixel. {\sc
  Commander} simultaneously estimates the angular power spectrum of the CMB
also which we will not be concerned about in this paper. We however still need to produce high quality maps if we want to look for
{cosmological information that go beyond the isotropic power spectrum.}

Our approach is
straightforward least-squares parameter fitting. This has been attempted
before using non-linear optimization algorithms \cite{brandt1994}. The
non-linear optimization algorithms however converge quite slowly and often
to the local minimum which may be far from the solution we are interested
in. We develop a new algorithm taking advantage of the fact that  most
of the parameters in the foreground model are linear, for example, the
amplitudes of different components. Even the non-linear parameters lie
within a narrow range. For example, the low frequency foregrounds originate
in free-free and synchrotron emission and are expected to have spectral
indices in the range $-2$ to $-4$. Similarly the spectral index of the dust
component is expected to lie between $2$ and $3$ (see the next section for the
exact definitions of spectral indices and the foreground model) and its temperature between
$\sim 10$ and $30~$K. This suggests that we can Taylor expand the foreground
model around a reasonable guess and solve the resulting linear
problem. Since the Taylor series is a good approximation within a narrow
range around the
central value, we expect the linear approximation to be a very good
one. This forms the basis of our algorithm presented in the next section. 
In section \ref{lil} we present our algorithm explaining the reasoning
behind the different steps. In section \ref{ffp6} we apply our algorithm to
the publicly available simulations of the Planck sky model \citep{ffp62013} called FFP6
simulations\footnote{\url{http://wiki.cosmos.esa.int/planckpla/index.php/Simulation_data}}. In section \ref{co} we discuss  a
further refinement/extension of our algorithm by including model
selection. {We work with the simulations in most of the paper
  except in section
\ref{planckres} where we apply our algorithm to the publicly available Planck maps
at full resolution and present the resulting CMB and foreground component
maps.} Our results as well as the FORTRAN code will be made publicly
available at \url{http://www.mpa-garching.mpg.de/~khatri/lilcmb}.

\section{Linearized iterative least-squares parameter fitting
  ({\sc LIL})}\label{lil}
For definiteness we will work with Planck experiment. However, the algorithm
is quite general and is applicable to future experiments with many more
frequency channels compared to Planck \citep{core,pixie,prism} and also to more complicated
foreground models.
Following \citet{planckcomp} we fit the following 6-parameter model to the
7 Planck frequency channels from 30 GHz to 353 GHz. In the following we will refer to the
observed value of data = signal + noise in a particular pixel ($p$) in
frequency channel $j$ as $d_j(p)$ and we
number the frequency channels from lowest to highest i.e. 
$j\in \{1,\ldots,7\}$ for $\nu\in \{30,\ldots,353\}~{\rm GHz}$ respectively. Our parametric model is

\begin{align}
s_{\nu}(p)=&\Acmb(p)+\flf\Alf(p)\left(\frac{\nu}{\nulf}\right)^{\blf(p)}+\fco\Aco(p)\nonumber\\
&+\fdst\Adst(p)\frac{1}{\exp\left(\frac{h\nu}{\kB\Tdst}\right)-1}\left(\frac{\nu}{\nudst}\right)^{\bdst(p)},\label{Eq:model}
\end{align}
where $p$ is the pixel number in
HEALPix\footnote{\url{http://healpix.sourceforge.net}} nested numbering
scheme \citep{healpix}, $s_{\nu}$ is the total sky emission in units of
$K_{\rm CMB}$ at frequency $\nu$, $A_i(p)$ is the amplitude of the component $i\in \{{\rm
  CMB,co,lf,d}\}$ where the abbreviations are for the CMB, CO line
emission, low frequency emission including synchrotron, free-free and
anomalous microwave emission, and dust emission respectively. The factors
of $f_{\nu}^i$ convert the Rayleigh Jeans temperature (or ${\rm K}_{\rm RJ}~{\rm Km/s }$ for CO) to the thermodynamic
CMB units, $K_{\rm CMB}$, and also include the color correction factors,
$\nu_0^i$ \citep{lfical,hfical} are the reference frequencies with
$\nulf=30~{\rm GHz}$ and $\nudst=353~{\rm GHz}$ and $\beta_i$ are the
spectral indices. The $\fco$ in addition includes relative amplitudes of
the 
CO lines in units of ${\rm K}_{\rm RJ}~{\rm Km/s }$ in different frequency channels which we keep fixed following
\citet{planckcomp} as (1:0.595:0.297) for 100 GHz , 217 GHz and 353 GHz
channels respectively and zero for all other channels.  We will assume the dust temperature to be constant, $\Tdst=18
~{\rm K}$. In principle it is possible to include higher frequency channels
at 545 GHz and 857 GHz and allow the dust temperature to vary. However, it is
not expected to lead to a significant improvement in the foreground
separation as far as CMB is concerned since the additional information
provided by the higher frequency channels is absorbed in the additional
complexity of the model as well as having to apply the simple dust emission
model over a wider range of frequencies than is physically justified. Note
that our definitions are slightly different from \citep{planckcomp} to make
the formulae a little simpler. We will also suppress the
  argument $p$ in the following since we work with one pixel at a time, so
  the value of data in $j^{\rm th}$ frequency map is denoted by just $d_j$.

We want to do a Taylor series expansion of Eq. \ref{Eq:model} around some
initial guess for the parameters. Therefore the zeroth step for our
algorithm is the following:
\begin{enumerate}[label=(\arabic*),start=0]
\item Set the initial values of parameters for the current pixel as
  follows: $\Acmb=\min(d_4,d_5,d_6)$, $\Alf=(d_1-\Acmb)/f^{\rm lf}_{30~{\rm GHz}}$,
  $\Adst=(d_7-\Acmb)/f^{\rm d}_{353~{\rm GHz}}$, where we ignore the color
    correction in $\flf$ and $\fdst$ which then just convert from
    Rayleigh-Jeans units to the thermodynamic units. Similarly using
    $d_1-\Acmb$ and $d_2-\Acmb$ as estimates of low frequency foreground in
    the lowest two channels we can solve for $\blf$ and using $d_7-\Acmb$
    and $d_6-\Acmb$ as estimates of dust foregrounds we can solve for the
    initial $\bdst$. Note that we do not need initial guess for CO
    amplitude since it is linear and sub-dominant compared to the other components.
\item Do Taylor expansion around the current value of the parameters
  (or the initial guess if first iteration). 
\begin{align}
  s_{\nu}\approx s'_{\nu}=&
  \Acmb+\fco\Aco+\flf\Alf\left(\frac{\nu}{\nulf}\right)^{\blf}\left[\dellf+\delblf\blf\ln\left(\frac{\nu}{\nulf}\right)\right]\nonumber\\
&+\fdst\Adst\frac{1}{\exp\left(\frac{h\nu}{\kB\Tdst}\right)-1}\left(\frac{\nu}{\nudst}\right)^{\bdst}\left[\deldst+\delbdst\bdst\ln\left(\frac{\nu}{\nudst}\right)\right]\label{Eq:taylor}
\end{align}
In the above equation the $\Alf,\blf,\Adst,\bdst$ are fixed from previous
iteration.
We thus have a 6 parameter linear model with the parameter vector
$x=(\Acmb,\Aco,\dellf,\delblf,\deldst,\delbdst)$. We have defined $\dellf$
and $\deldst$ as multiplicative corrections to the amplitude while the
spectral indices $\delblf,\delbdst$ are fractional corrections so that the
actual indices are $\blf(1+\delblf), \bdst(1+\delbdst)$ and we expect that
$\delbdst,\delblf\lesssim 1$ so that the Taylor series is a good
approximation.
\item We now solve the linear least squares problem minimizing
\begin{align}
\chi^2=\sum_i\left(\frac{s'_{\nu_i}-d_i}{\sigma_i}\right)^2 \equiv \sum_i \left[(Mx)_i-d'_i\right]^2,
\end{align}
where $d'_i\equiv d_i/\sigma_i$ and $\sigma_i$ is the standard deviation of noise
in channel $i$ in the current pixel $p$. The problem is easily solved by
numerous linear algebra techniques. We use LQ/QR factorization routines of the
Intel Math Kernel Library. The solution of the least squares problem gives
us a direction to move our current parameter vector as well as the
amplitude of the step which is just the actual least squares solution. In
practice we may decide not to take the full step but only move a fraction
of the amplitude in the relevant direction. We will come back to this point
below.
\end{enumerate}

So far we have done nothing new. The steps (1) and (2) are in fact just the
standard Taylor expansion to linear (or quadratic) order and the Gauss-Newton
step respectively. These two steps, with or without some
modification,  form the basis of most non-linear optimization algorithms
(\citealp[see e.g.][]{gillmurray,nr}; \citealp{tr}). From
Eq. \ref{Eq:taylor} it is clear why we do not arrive at the correct
solution in just one step. Although this linear model is a good
approximation, in fact the term multiplying the amplitude parameters  $\dellf$ and $\deldst$ have
the wrong spectral indices since they also change in the step from their
current values and vice-versa for the terms multiplying the spectral index
parameters $\delblf$ and
$\delbdst$. This is in fact the only place the non-linearity of our model
manifests itself. This also suggests a cure. 
\begin{enumerate}[label=(\arabic*),start=3]
\item Use the results of step (2)
only to update the spectral indices $\blf\rightarrow\blf(1+\delblf)$ and
$\bdst\rightarrow\bdst(1+\delbdst)$. Fix the spectral indices and
repeat step (2) for the smaller parameter vector
$x'=(\Acmb,\Aco,\dellf,\deldst)$. Update the amplitudes with the new solution.
\end{enumerate}

With the spectral indices fixed the problem is in fact linear and can be
solved exactly. Thus we arrive at the minimum in a subspace of our
full parameter space. 

\begin{enumerate}[label=(\arabic*),start=4]
\item  Exit if we
  meet one of the exit criterion given below otherwise go back to step (1)
\end{enumerate}

We are therefore searching for a minimum in the two non-linear directions,
 which are almost uncorrelated with each other.
while always remaining at the global minimum in the  subspace, defined by
the component amplitudes, of the full
parameter space.

Finally we can also improve our initial  guess and make the initial amplitudes and
indices of the model compatible with each other by solving the
least-squares problem in the subspace of amplitudes ($x'$) but keeping the
initial guess for the indices.

\begin{enumerate}[label=(\arabic*$'$),start=0]
\item  After step (0) solve the least squares problem for the amplitudes
  ($x'$) similar to step (3) and update the amplitudes before proceeding to
  step (1).
\end{enumerate}

It turns out that the combination of the linearized model,
Eq. \ref{Eq:taylor}, combined with additional sub-iteration step (3) makes
the algorithm extremely efficient in finding the minimum of the
$\chi^2$. We have in fact a significant advantage over  a general
non-linear problem. The linear model, Eq. \ref{Eq:taylor}, is a good
approximation over almost the entire range of allowed parameter space since
the spectral indices cannot vary by a large amount and therefore our
minimum lies within the range of validity of the expansion. This is not true in
general and the Taylor expansion at linear or quadratic order for a
general 
non-linear problem would probably be valid in a small region of parameter
space which may not include the minimum we are after.
It is of course very difficult to prove that we have indeed found
the global minimum and the only way to test the quality of results is to
use simulations which we do in section \ref{ffp6}.
\begin{figure*}
\resizebox{\hsize}{!}{\includegraphics{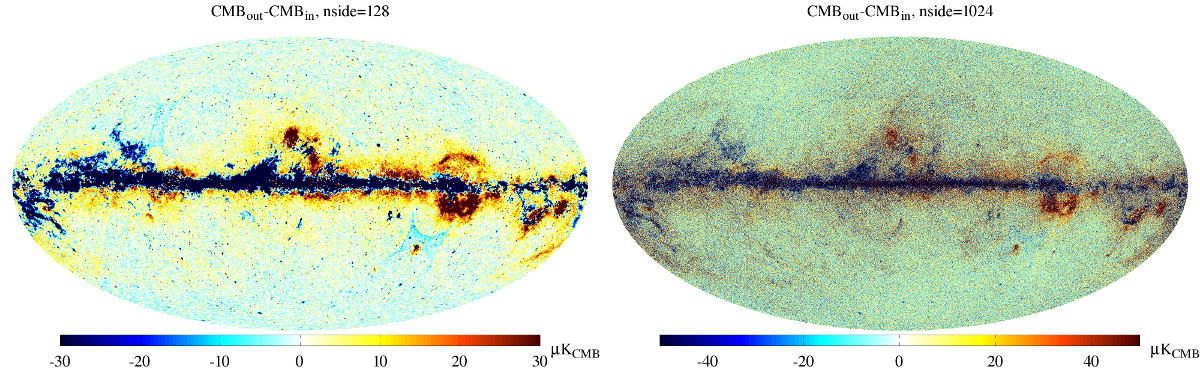}}
\caption{\label{Fig:rescmb} CMB Residuals for the FFP6 simulation, (${\rm CMB}_{\rm
    out}$-${\rm CMB}_{\rm in}$). The residuals are less than few $\mu{\rm
    K}$ at high latitudes at nside=128. There is also no visible bias at
  high latitudes and the residuals are consistent with noise. Note that our color scale is different
  from that used by \citep{planckcomp}.  A monopole and dipole calculated at latitudes
  $|b|>30^{\circ}$ has been subtracted from the residual maps.}
\end{figure*}
\begin{figure*}
\resizebox{\hsize}{!}{\includegraphics{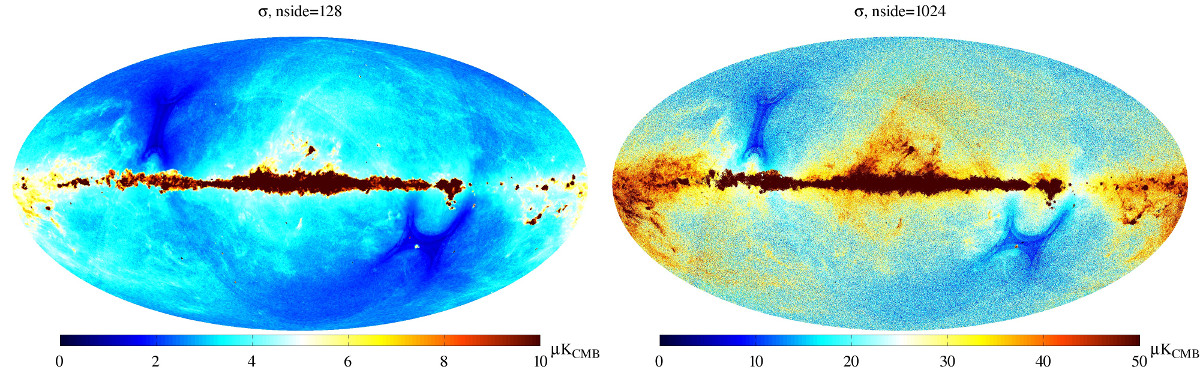}}
\caption{\label{Fig:err} The standard deviation marginalised over the
  foreground components. The errors roughly follow the pattern of
  foregrounds on the sky correctly modelling the uncertainties in
  foreground subtraction.}
\end{figure*}
\begin{figure*}
\resizebox{\hsize}{!}{\includegraphics{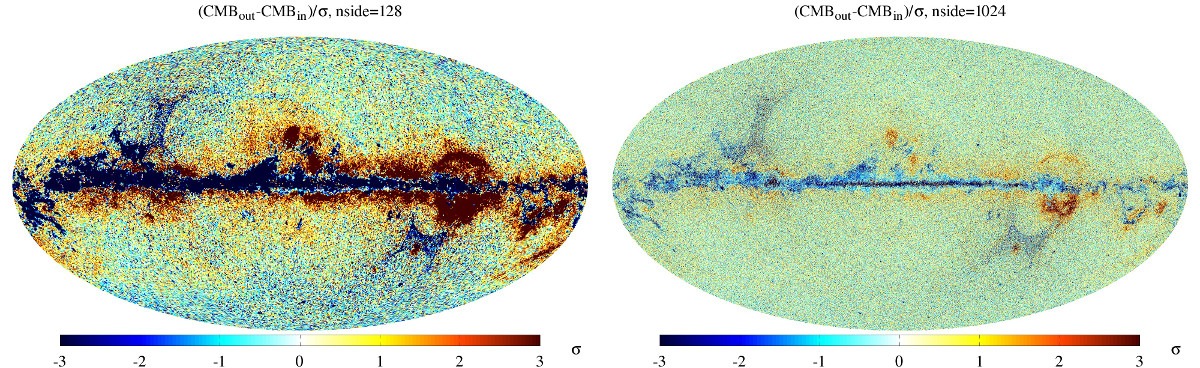}}
\caption{\label{Fig:std} The normalized error defined as ratio of
  residuals to the standard deviation estimate. If the error estimates are
  correct then this map should resemble Gaussian noise. The features at
  the Ecliptic poles are better visible in this map compared to the Fig. 2 owing to extremely low
  noise and shows that there are systematic errors not included in our
  error estimates.}
\end{figure*}

\subsection{Additional constraints and exit criterion}
We need to add additional constraints to the above algorithm motivated by
the physical nature of the foregrounds. In other words we want to solve the
least-squares problem subject to certain constraints so that we remain in
the physically relevant parameter space in the presence of noise.
\begin{enumerate}
\item We want to restrict the allowed range of the spectral indices
  $\blf,\bdst$. We impose a somewhat loose bound in the regions where
  foregrounds have high amplitude and a tighter bound in the regions
  with low foregrounds. This is an attempt to emulate a prior in the
  Bayesian sense.  If a foreground component is detected with high
  signal to noise the bounds  have no effect and we can allow a wider
  freedom for the spectral index to vary. The  bounds we use are 
  as follows:
\begin{align}
d_1-d_2 > 10^{-3}~{\rm K_{\rm CMB}} & \rightarrow -5\le \blf \le -1 \nonumber\\
d_1-d_2 \le 10^{-3}~{\rm K_{\rm CMB}} & \rightarrow -3.5\le \blf \le -2.5 \nonumber\\
d_7-d_6 > 10^{-2}~{\rm K_{\rm CMB}} & \rightarrow 1\le \bdst \le 5 \nonumber\\
d_7-d_6 \le 10^{-2}~{\rm K_{\rm CMB}} & \rightarrow 2\le \bdst \le 3 
\end{align}
It may happen, usually where foregrounds have too small S/N, that there is
no local minimum in the direction of spectral indices, where the derivative of
$\chi^2$ vanishes, within the bounds.  We will still have a minimum value for the $\chi^2$ within
the bounds which we can use as the best value of our parameters, usually at
the one of the boundaries. In that
case it is not possible to use the Fisher matrix to estimate the
uncertainties and in particular  it will in general not be positive
definite. In such situations, since foregrounds are anyway small and
therefore their influence on the errors on the CMB component should be
minimal, we calculate the covariance after removing the columns/rows
corresponding to the non-linear parameter (spectral index) which hit the
boundary or which is causing the Fisher matrix to be non-positive definite.

\item In the regions of low or non-existent foregrounds there are less
  components in reality than present in the model. Therefore we will be
  fitting most of the parameters to noise. Also the lowest frequency
  channels in Planck have much higher noise and poorer resolution and they
  may pull down the S/N of the final component maps. To avoid this we
  explicitly fit the low frequency foregrounds only to the lowest 4
  frequency channels and dust to the highest 4 frequency channels. We do
  this by setting following conditions
\begin{align}
d_1-d_2\le 10^{-3}~{\rm K}_{\rm CMB}&\rightarrow\flf|_{\nu\ge 143~{\rm GHz}}=0\nonumber\\
d_7-d_6\le 10^{-2}~{\rm K}_{\rm CMB}&\rightarrow\fdst|_{\nu\le 70~{\rm GHz}}=0 
\end{align}

\item It is possible that the amplitudes would tend to go negative during
  the iterations. We take this also as a sign that the foregrounds are
  negligible and are just being fitted to noise. When this happens for the
  dust and low frequency amplitudes, we multiply the existing amplitude by
  a factor of 0.1 instead of updating to negative value. For the CO
  amplitude, if the CO contribution to a channel  goes below $1\%$
  of noise level in all of the three channels in which the CO contribution is non-zero, we remove CO as a
  foregrounds component 
  for the next 5 iterations. If the final parameters with minimum $\chi^2$
  were fitted without CO then we remove the corresponding columns/rows also from
  the Fisher matrix.
\end{enumerate}

Finally we have the following exit criterion out of the iteration loop:
\begin{enumerate}
\item Exit if $\chi^2<0.1$ or change in $\chi^2$ in the previous 2 iterations
  is less than $10^{-3}$. 
\item The number of iterations has exceeded 100. 
\end{enumerate}
On exit the output parameters correspond to the iteration which had the
minimum $\chi^2$ among all the iterations. This is necessary since as we
mentioned earlier, the global minimum of the non-linear problem is not
necessarily the physical solution that we desire if it lies outside the
physical constraints imposed by us. Within the physical constraints, the
minimum $\chi^2$ encountered may be a saddle point or may even have
non-vanishing first derivative. Therefore the final converged value of the $\chi^2$ may not be a
minimum but an asymptote flying off to outside the allowed parameter region
or even to infinity. Our prescription gives us
the best values of parameters within the physical constraints imposed by
us.

\section{Validation of the algorithm on the Planck Sky Model FFP6
  simulations}\label{ffp6}
We know apply our algorithm to the publicly available FFP6 simulations
of the Planck sky model (PSM) \citep{ffp62013}. In particular we use the nominal survey full sky
signal maps. These maps include estimates of the variance in each pixel
which we use as estimates of the noise variance in our least squares
fitting. The low frequency instrument (LFI) maps in the lowest three
channels are at lower resolution of HEALPix nside
1024 compared to the nside 2048 for the high frequency instrument (HFI). We
upgrade the LFI maps to 2048 also scaling the variance. Since the LFI maps
are anyway more noisy compared to the HFI maps, this upgradation does not
affect the CMB or the high frequency components. To get the best low
frequency components the component separation can be repeated by degrading
all maps to, for example,  nside=512. 

 Since Planck measures only the change in
signal across the sky and not the absolute sky brightness, there are
relative offsets between different channels. Therefore  estimates of monopole and dipole should be subtracted from the maps
to prevent systematic errors. In the FFP6 simulations we find that the
offsets are small enough that they do not matter. For the actual Planck
data we will however subtract the best fit values of the monopoles and
dipoles provided by \citet{planckcomp}.

The residual map, i.e. difference (${\rm CMB}_{\rm out}$-${\rm CMB}_{\rm in}$) between the  output
from LIL (${\rm CMB}_{\rm out}$) and the input CMB map in FFP6
simulation (${\rm CMB}_{\rm in}$) is shown in Figure \ref{Fig:rescmb}. We
show the differences for two HEALPix resolution at nside 1024 and 128. 
The residual at nside 128 can be compared to that from the other algorithms
used by the Planck collaboration (Fig. 7 in \citet{planckcomp}). The
residuals are about the same level as other algorithms and of order of few
$\mu {\rm K}$ at high latitudes. There is a systematic feature around the
Ecliptic poles where because of the Planck scanning strategy the depth of
the survey and the noise
levels change abruptly. We discuss this and the small systematic  bias near
the galactic plane where the residuals are preferentially positive in
detail section \ref{co}. We note that similar systematic effects and biases
are also
present in the Planck collaboration analysis of FFP6 simulations \citep{planckcomp}. In
particular the feature near the Ecliptic poles can be seen in the {\sc Commander}
residual maps.
\begin{figure}
\resizebox{\hsize}{!}{\includegraphics{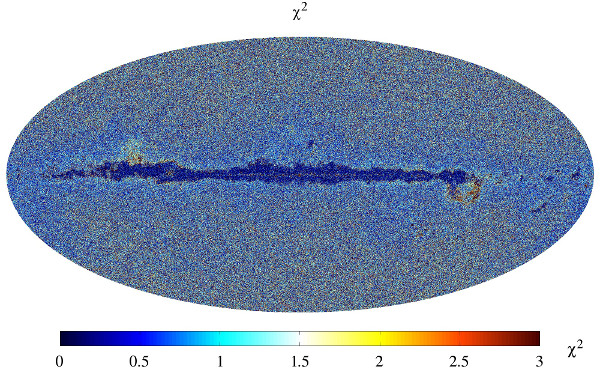}}
\caption{\label{Fig:chisq} The  $\chi^2$ map for the least squares fit performed
  by LIL  in FFP6 simulations. For 1 or 2 degrees of freedom that we have with or without the CO
component, the average $\chi^2$ is expected to be between 1 and 2.}
\end{figure}
\begin{figure}
\resizebox{\hsize}{!}{\includegraphics{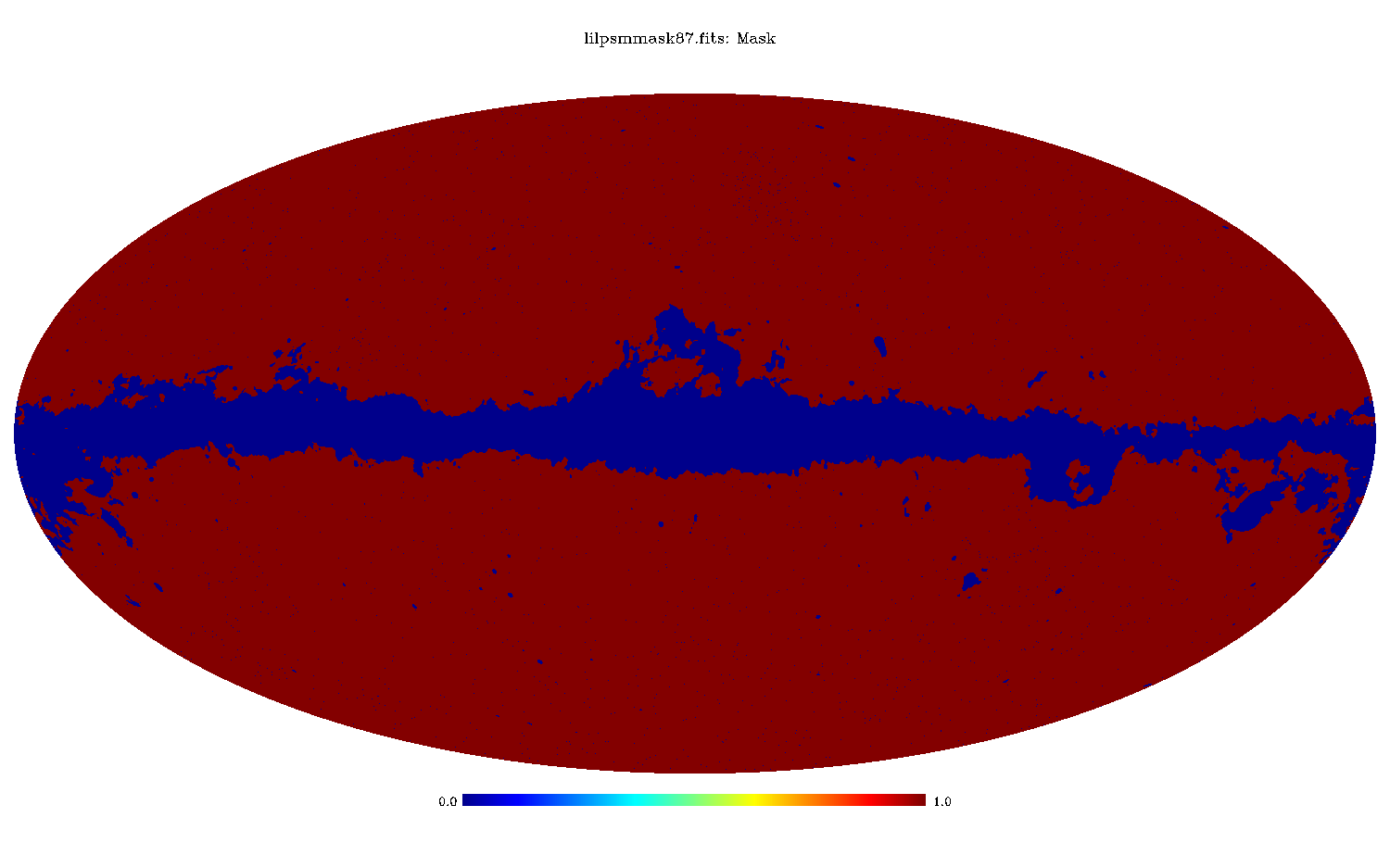}}
\caption{\label{Fig:mask83} The  mask used in further validation of the
  results from FFP6 simulation yielding a sky fraction usable for CMB of $86.4\%$.}
\end{figure}

\subsection{Error estimation}

We estimate errors on the final components using Fisher matrix. As
mentioner earlier, for pixels to which the model without the CO component
was fitted we delete the corresponding columns/rows from the Fisher matrix. Also
for pixels which did not converge to a minimum in the direction of one or
both of the spectral indices those columns/rows are also deleted. We will see
that this prescription gives us a reasonable estimate for the errors. The
Fisher matrix is calculated at full resolution, nside=2048. For lower
resolutions we  combine the errors in the smaller pixels in each
larger pixel assuming that the errors/noise are uncorrelated by first
degrading the variance map to new nside, which averages the variance, and then dividing
the variance in each pixel by ${\rm npix}_{2048}/{\rm npix}_{\rm
  nside}=2048^2/{\rm nside^2}$,
where ${\rm npix}_{\rm nside}\propto {\rm nside^2}$  is the total number of
pixels in the map. This is an
approximation since the beams of all frequency channels are larger than the
size of any pixel. We show the standard deviation error and normalized error map in Figures
\ref{Fig:err} and \ref{Fig:std}. The normalized error map ($({\rm CMB}_{\rm
    out}$-${\rm CMB}_{\rm in})/\sigma$) is just the
ratio of the residual map to the standard deviation  ($\sigma(p)$) map.
\begin{figure*}
\resizebox{\hsize}{!}{\includegraphics{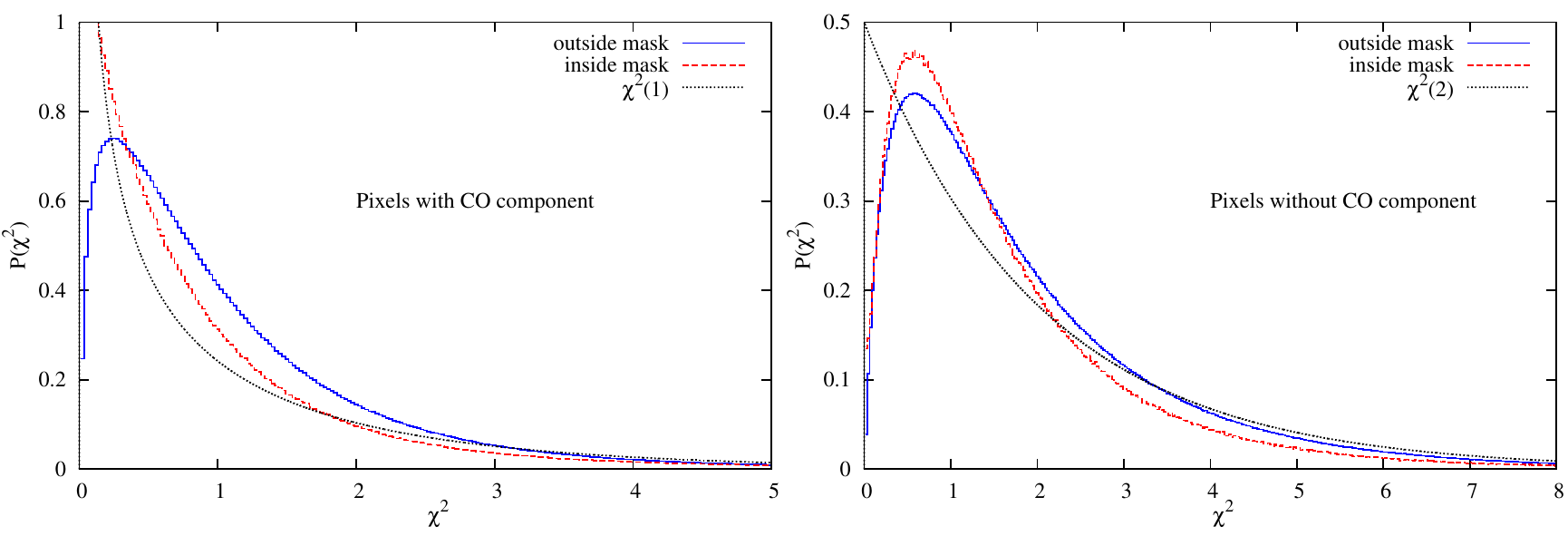}}
\caption{\label{Fig:psmchi} The $\chi^2$ probability density function
  obtained by LIL  for FFP6 simulations and compared with the theoretically expected $\chi^2$
  distributions with 1 and 2 degrees of freedom. The LIL distribution
  matches well with the theoretical curves for
  large $\chi^2$ values and inside the mask in the left figure where all the modelled components are present.}
\end{figure*}

The standard deviation  $\sigma$ roughly follows the distribution of
foregrounds on the sky. This is because there are uncertainties associated
with the foreground subtraction and which we marginalise over. The
normalized error is an estimate of how good and compatible our CMB signal
and error estimates are. For perfect reconstruction of the mean signal and
errors this map should resemble Gaussian noise. We see that this is so
except in the galactic plane where our simple model does not capture all
the complexities of the foregrounds. We have also neglected the noise
correlations between pixels and therefore in the degraded maps the noise is
slightly underestimated. In particular the features near the Ecliptic poles
are more prominent in the normalized errors map indicating that there are
systematic uncertainties in these extremely low noise regions which are not
included in our error estimates. The goodness of fit can be quantified by
the $\chi^2$, and the $\chi^2$ map is shown in Fig. \ref{Fig:chisq}. This
can be compared with the $\chi^2$ map of  {\sc Commander}
\citep{planckcomp}. One difference is that we get low $\chi^2$ values in
the galactic plane in contrast to {\sc Commander}, implying that our model
is a good fit to the data in the galactic plane. This difference from {\sc
  Commander} could be because of the tight prior they impose on the
spectral indices while we allow a wider range of spectral indices in the
high foreground regions. {We should emphasize here that our
  $\chi^2$ and that of {\sc Commander} are not equivalent. Our $\chi^2$
  is calculated at the  best fit parameter values or the likelihood maximum
  whereas  that of
  {\sc Commander} are calculated for samples from the posterior.}

{The value of $\chi^2$ is determined by
  many factors. The obvious one is the degrees of freedom or the difference
  between the number of data points and number of parameters. Normally we
  expect the $\chi^2$ to be distributed as a $\chi^2$-distribution with the
  average value equal to the number of degrees of freedom. However this is
  only true if we allow complete freedom for parameters to take on any
  value. Addition of constraints increases the degrees of freedom by
  reducing the effective number of parameters, in our case the non-linear
  spectral indices, which are not free to vary outside the constraint
  imposed by us. The best fit solution is not the maximum of the likelihood
  in the full 6-dimensional parameter space. As we move away from the
  galactic plane, therefore, we expect the $\chi^2$ to move towards an
  average value of $>1$ as  first the low frequency index and then the dust
  index hits the constraints and the corresponding amplitudes go towards zero. The foregrounds signal
  becomes too weak to
  find the best fit value (where the likelihood has a maximum) of  the
  non-linear parameters within the constraints and value of $\chi^2$ increases. In the galactic plane there is enough signal
  to noise that the non-linear parameters can be constrained and the
  effective degrees of freedom returns to one. In the very center the
  foregrounds become too complex and the $\chi^2$ increases again, this
  time because our model is too simple to account for all the signal even
  in the FFP6 simulations. For example, at low frequencies, when both the
  synchrotron and spinning dust emission are present, the spectrum which cannot be
  fit by a power law and we see regions of high $\chi^2$ at these places on
  the map.
}
\subsection{Mask}\label{mask}

To better analyze the quality of our component separation for the CMB as
well as the foreground components we need to mask out the worst regions of
the sky. We use the dust amplitude at 353 GHz, low frequency amplitude
at 30 GHz and the standard deviation maps as estimated by LIL to construct
the mask. We smooth the maps with a 30 arcmin full width half maximum (FWHM) Gaussian beam and threshold the dust amplitude at 2 MJ/Sr, low frequency
amplitude at $600 ~\mu{\rm K}$ and standard deviation estimate at $75
~\mu{\rm K}$ resulting in masking $13.6\%$ of the sky. We also mask pixels
with the $\chi^2\ge 10$. This scheme also
masks the brightest point sources as can be seen in Fig. \ref{Fig:mask83} and we do
not use any additional point source mask for the results from FFP6 simulations.

\subsection{Validation}

If our model was a good description of the Planck sky then the $\chi^2$
will follow the $\chi^2$-distribution with 1 or 2 degrees for freedom for
the pixels with and without the CO components respectively. The probability
density function of $\chi^2$ is shown in
Fig. \ref{Fig:psmchi} for the pixels fitted with
and without the CO component and compared respectively along with the
theoretically expected $\chi^2$-distributions with 1 and 2 degrees of
freedom labeled $\chi^2(1),\chi^2(2)$. LIL distributions match the
theoretical curves quite well despite the fact that our foreground model is
much simpler compared to the actual foregrounds. The deviations we see are
similar in nature to those observed in the Bayesian codes
\citep{eriksen2006}. 
\begin{figure}
\resizebox{\hsize}{!}{\includegraphics{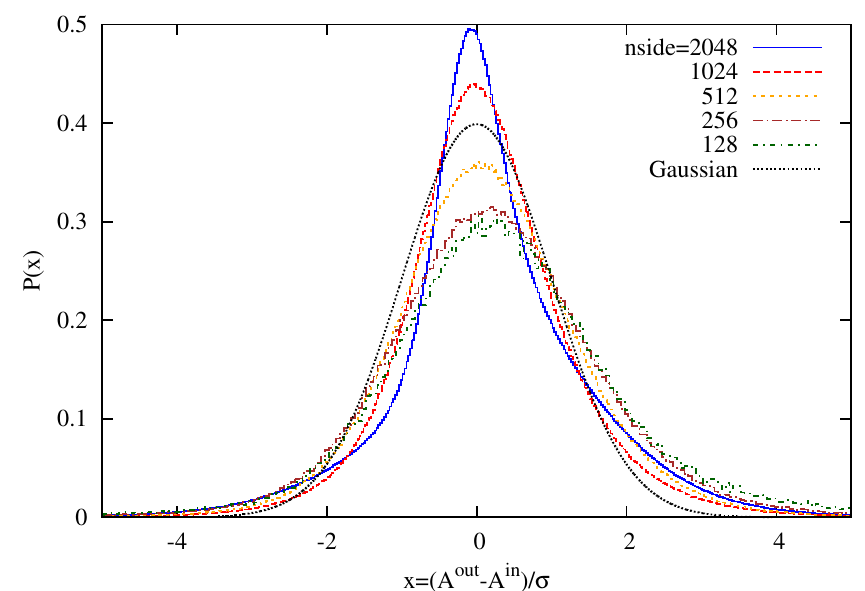}}
\caption{\label{Fig:pdfcmb} The  probability density function of normalized
  error (Fig. \ref{Fig:std}) $(A_{\rm CMB}^{\rm in}-A_{\rm CMB}^{\rm
    out})/\sigma_{\rm CMB}$ for different resolutions {for pixels outside
  the mask.} For high resolution
  the curves are close to the expected Gaussian but the error is
  underestimated as we go to the lower resolutions because we ignored the
  correlations in the noise. We however note that the distribution
  approaches a Gaussian as the average pixel size approaches the resolution
  of the highest frequency channels. For the foregrounds results therefore we will
  use nside=1024 for dust and nside=512 for the low frequency component. 
}
\end{figure}

The $\chi^2$ on average is a little
larger than the theoretical values. We should however expect this since not
all the components fitted in our model are present everywhere in the
sky. The obvious example is the CO component which is present at detectable
levels only near the galactic plane. Even the low frequency and dust
components are not present everywhere on the sky. The low frequency
components are below the noise levels in a major fraction of the sky. The
dust component also is very weak  in the 217 GHz and lower frequency
channels in a good fraction of the sky and so the spectral index of even
the dust component cannot
always be determined with any accuracy. {The number of degrees of freedom at
therefore $>1$, as discussed above,} and this shows up in the
higher values of $\chi^2$. Our hypothesis is supported by the fact that the
$\chi^2$ distribution inside the mask for pixels where there is detectable
CO is much closer to the theoretical curve. This reasoning also explains
why the algorithms such as SMICA, NILC and SEVEM, which have an equivalent implicit
model with smaller number of parameters, do so well. The above arguments  also suggest
that  fitting the most general model to the data may not be the optimal approach. 
What we should really do it fit  many models to the data and choose from
the models the one which fits best given the number of parameter and the degrees
of freedom. We will do exactly this for the CO component in section
\ref{co}. 

\begin{figure*}
\resizebox{\hsize}{!}{\includegraphics{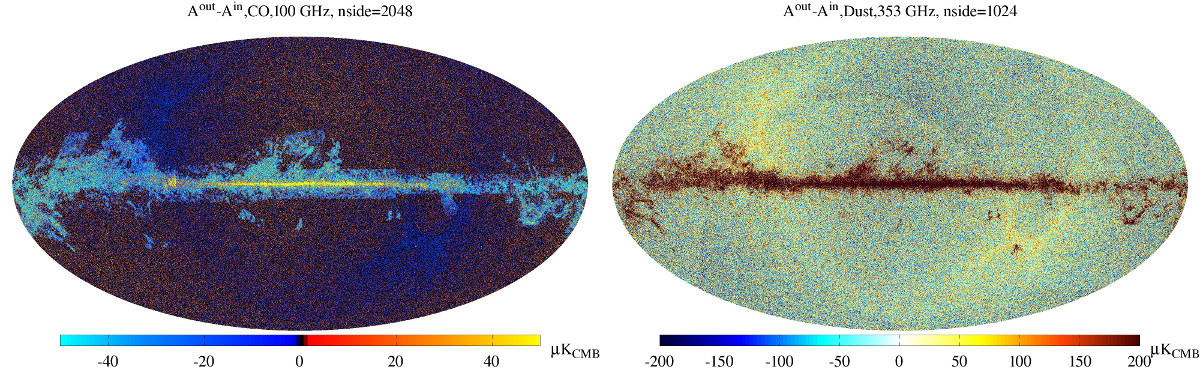}}
\caption{\label{Fig:hfires} Residuals in the CO and dust components. The CO
  residual away from the galactic plane follows the hit count map decided
  by the Planck
  scanning strategy. A monopole and dipole calculated at latitudes
  $|b|>30^{\circ}$ has been subtracted from the residual maps. Note that we
  are using a different color scheme for the CO to better highlight the
  non-zero pixels.
}
\end{figure*}

\begin{figure*}
\resizebox{\hsize}{!}{\includegraphics{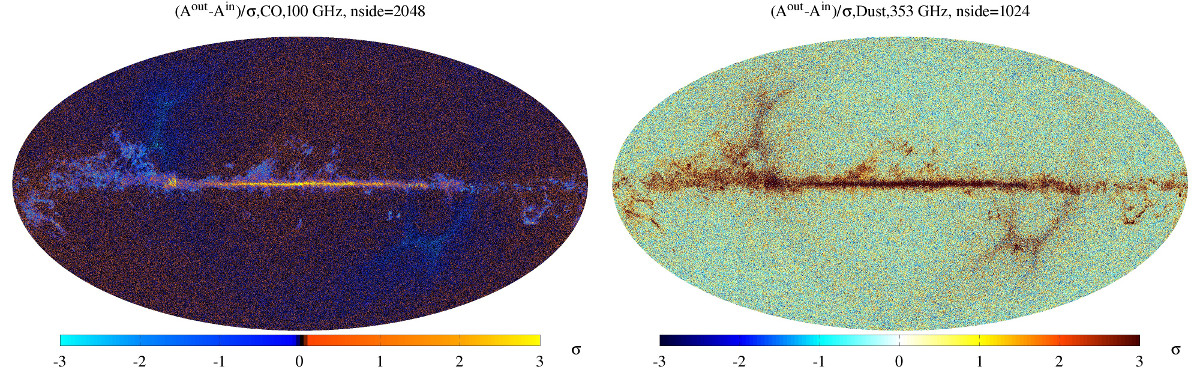}}
\caption{\label{Fig:hfistd} The normalized errors in the CO and dust
  components. The difference in the CO plot from Fig. \ref{Fig:hfires}
    around the Galactic plane is because in this plot we only include
    pixels for which we fitted for the CO component while
    Fig. \ref{Fig:hfires} includes all Pixels. In particular CO in the
    rectangular strips present in the FFP6 simulation is not detected as
    the signal is too low.
}
\end{figure*}

We show in Fig. \ref{Fig:pdfcmb} the probability distribution (PDF) of normalized
error for the CMB for different resolutions {outside the mask.} At high resolution the PDF is
close to a Gaussian. The deviations from the Gaussian are of similar
magnitude to that
of \citep{eriksen2006}. The reason for the under-estimation of errors as we
decrease the resolution is because we have calculated the errors ignoring the
noise correlations between pixels. The average size of the pixels is closer
to the actual resolution of the Planck HFI channels for nside = 512 and
1024 and these distributions are closest to the Gaussian. This fact gives
us some confidence that we are recovering correctly the best fit values as
well as the error bars.

\begin{figure*}
\resizebox{\hsize}{!}{\includegraphics{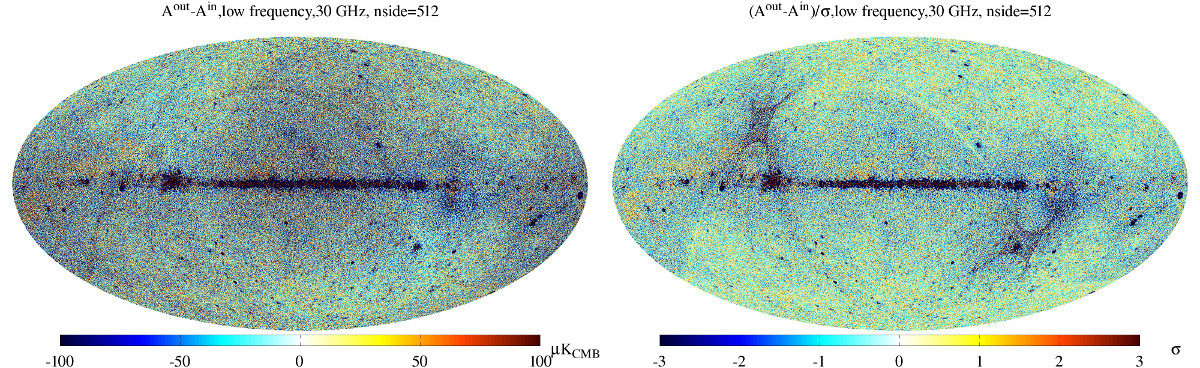}}
\caption{\label{Fig:lfi} The residual for the low frequency component at
  30 GHz and the normalized error. Away from the galactic plane the
  residuals and errors are consistent with Gaussian noise.
}
\end{figure*}
\begin{figure*}
\resizebox{\hsize}{!}{\includegraphics{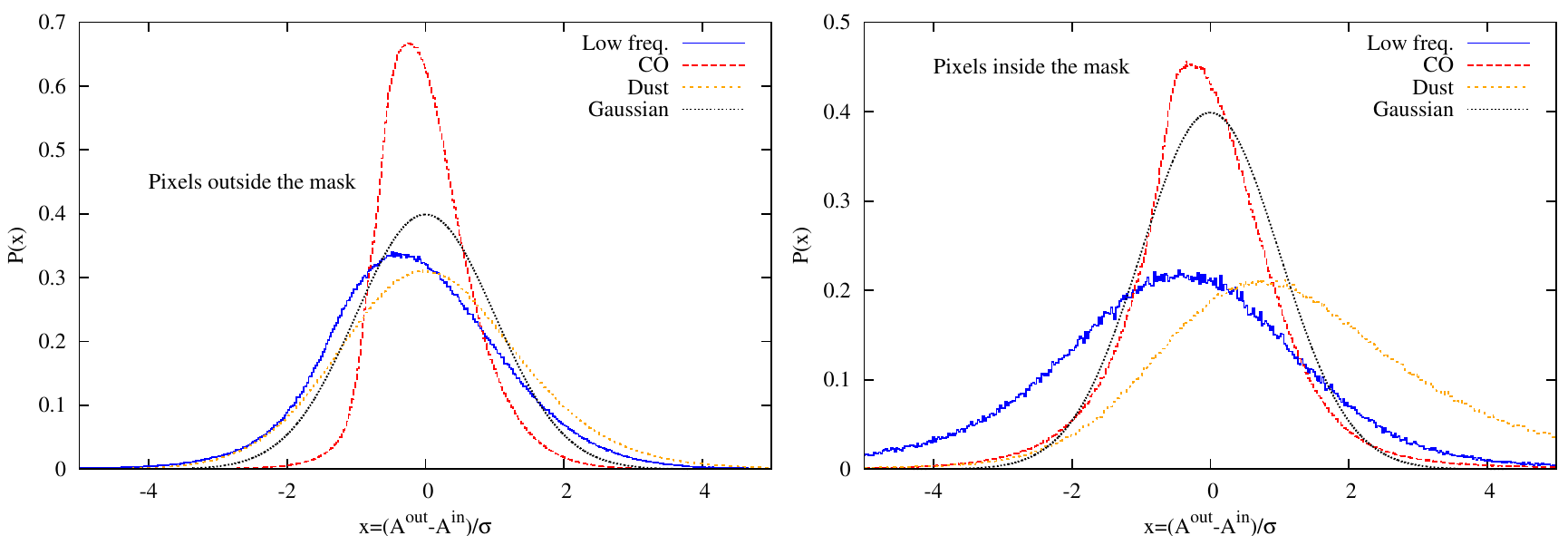}}
\caption{\label{Fig:pdffg} The  probability density function of normalized
  error (Figs. \ref{Fig:hfistd} and \ref{Fig:lfi}) $(A_{\rm in}-A_{\rm
    out})/\sigma$ for different foreground component amplitudes in our
  foreground model. Outside the mask, the distributions are closed to
  Gaussian except for the CO for which there is a significant overestimation
  of error. Note that the CO plots are for pixels at nside=2048 while
  the  plots for dust and low frequency components were made after
  degrading the residual and error maps to nside=1024 and 512 respectively.
}
\end{figure*}
\begin{figure*}
\resizebox{\hsize}{!}{\includegraphics{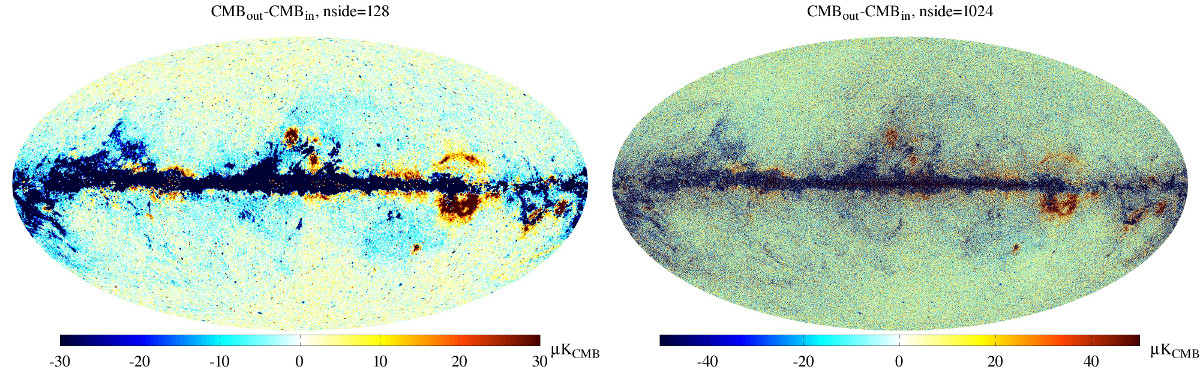}}
\caption{\label{Fig:rescmbco}CMB Residuals for the FFP6 simulation, (${\rm CMB}_{\rm
    out}$-${\rm CMB}_{\rm in}$) for LIL-MS. There
  is marked improvement compared to LIL near the galactic plane and the Ecliptic features
  are also weaker and less sharp.}
\end{figure*}

We show in Figs. \ref{Fig:hfires} and \ref{Fig:hfistd} the residuals and
the normalized error for the two high frequency foreground components, the
CO and the dust amplitudes, at the corresponding reference
frequencies. Figure \ref{Fig:lfi} shows the same quantities for the low
frequency component of our foreground model. Away from the Galactic plane, the residuals are consistent
with noise. The CO residuals follow the Planck hit count map and is
consistent with the CO component just fitting noise away from the galactic
plane. The true CO emission in the FFP6 simulation is zero away from the
galactic plane. Note that a large fraction of the pixels are fitted without the CO
component ($54\%$ pixels outside the mask). These pixels have the CO amplitude set to
zero and do not have an error on the CO amplitude. For the CO residual map we
have used all pixels while for the CO normalized error map only the pixels
with the CO component are non-zero. This accounts for the
difference in structure of the two maps since the residual map includes
differences for pixels which had a non-zero value in the original FFP6
simulation but are not detectable by LIL in the Planck data. These pixels
are mostly in form of the rectangular bands around the Galactic plane in
the residual map, Fig. \ref{Fig:hfires}.  We have also used a modified
color scheme to better highlight the non-zero pixels against the background
of pixels with CO amplitude fixed to zero. Also since most pixels with the
CO component away from
the galactic plane   are
surrounded by the pixels without the CO component, there is no consistent
way to degrade the resolution of the error map and for the CO component we
present results at the full resolution, nside=2048.

Figure \ref{Fig:pdffg} shows the PDF of normalized errors
on the foreground amplitudes within and outside our mask. Outside the mask
the PDF is close to Gaussian and we do not see as significant an
underestimation of errors for the dust component as reported in
\citet{planckcomp}. Note that our plots are at a higher resolution compared
to \citet{planckcomp}, nside=1024 for
the high frequency components and nside=512 for the low frequency
component. Inside the mask the deviations from the Gaussian are
more significant indicating more complex foregrounds compared to what we
have modelled. 

Finally to validate the full foreground model including the spectral
indices, we show the residuals for sum of all foregrounds in the 7 Planck
frequency channels used in our analysis in Figs. \ref{Fig:fgreshfi} and
\ref{Fig:fgreslfi} in the Appendix. The foreground residuals are very low for the 70 GHz
channel of Planck Low Frequency Instrument (LFI) and 100 GHz, 143 GHz and
217 GHz channels of the Planck High Frequency Instrument (HFI) thus
affirming the accuracy of spectral indices inferred by LIL.

\begin{figure*}
\resizebox{\hsize}{!}{\includegraphics{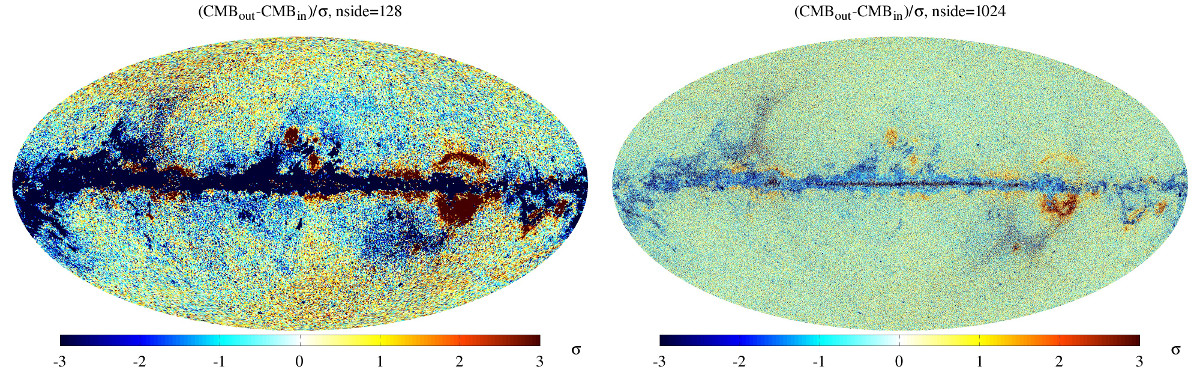}}
\caption{\label{Fig:stdco} The normalized error for LIL-MS.}
\end{figure*}
\begin{figure*}
\resizebox{\hsize}{!}{\includegraphics{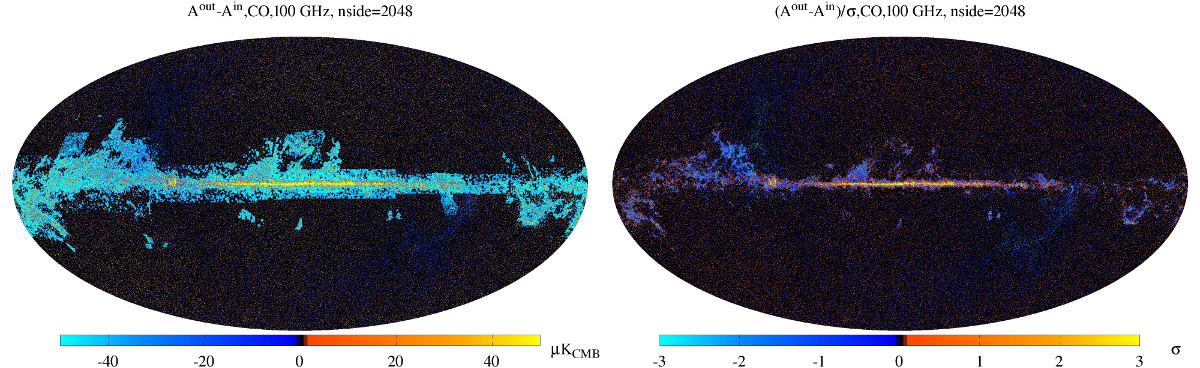}}
\caption{\label{Fig:coresstd} The residuals and normalized error for the CO
  component in LIL-MS. There is a marked
  improvement outside the Galaxy in the residuals compared to Figs. \ref{Fig:hfires}
  and \ref{Fig:hfistd}.}
\end{figure*}
\begin{figure*}
\resizebox{\hsize}{!}{\includegraphics{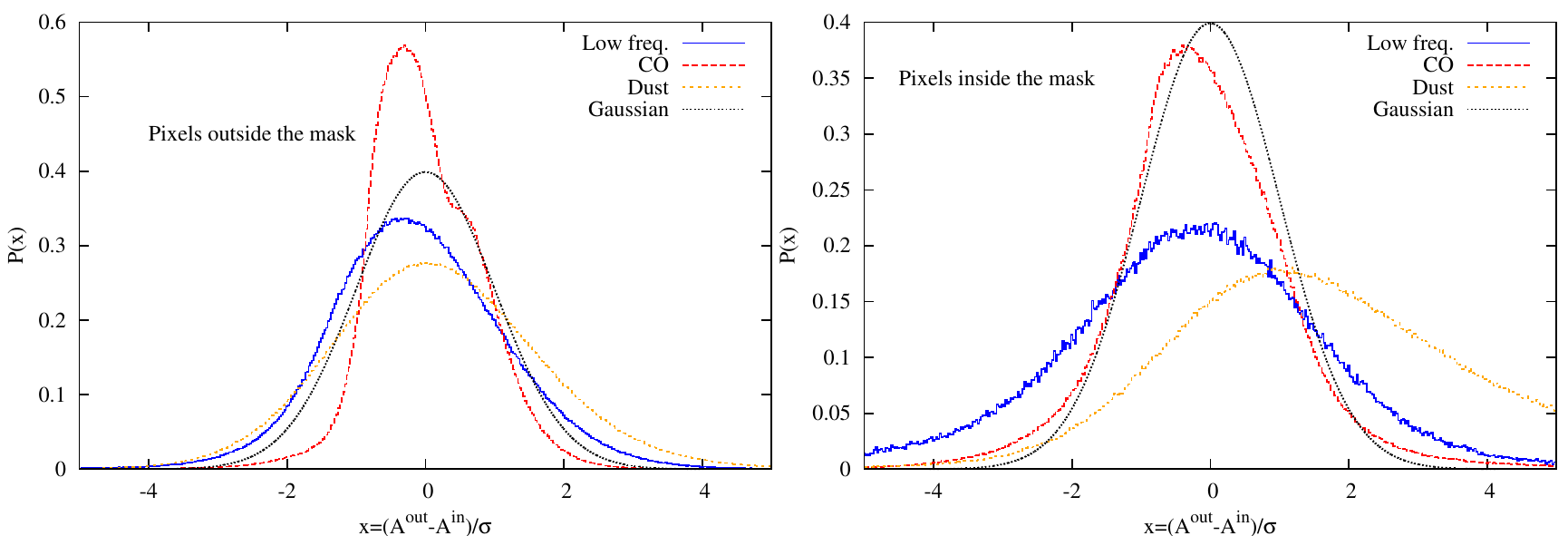}}
\caption{\label{Fig:fgpdfco} The  probability density function of normalized
  error $(A_{\rm in}-A_{\rm
    out})/\sigma$ for different foreground component amplitudes for LIL-MS. Outside the mask, the distributions are closed to
  Gaussian except for the CO for which there is a significant overestimation
  of error. There is however an improvement compared to the LIL. Note that the CO plots are for pixels at nside=2048 while
  the  plots for dust and low frequency components were made after
  degrading the residual and error maps to nside=1024 and 512 respectively.
}
\end{figure*}

\section{LIL-MS: Parameter fitting with model selection}\label{co}
We have so far emulated the approach of {\sc Commander} and shown that our
results are consistent. We can improve over this approach of fitting
parameters to a given model. One obvious improvement is suggested by
looking at Figs. \ref{Fig:hfires} and \ref{Fig:hfistd} for the CO
component. The CO component is detectable in the Planck data only in the
galactic plane and small regions around it, which show up as small scale
features in the maps. The large scale features are identical to the Planck hit
count maps and are the result of fitting the CO component to noise in most
of the sky where no CO is present. In LIL algorithm we did set the CO
component to zero in $51\%$ of all  pixels ($54\%$ if only considering
pixels outside the
mask) where the amplitude tended to go
negative. These pixels are evenly distributed over the sky and there are
still enough pixels with the CO component all over the sky that this does
not help with the large scale systematics.

Since the CO is present only in the high foreground regions which are
anyway masked for CMB analysis, the simplest solution to get the best CMB
maps would be to ignore the CO component altogether. This would give incorrect results
where the CO component is present but would not affect the cosmology if
these regions are masked. A more sophisticated approach, which does not
assume that CO is present only in the high foreground regions, would be to
fit models both  with and 
without the CO component and select the model which fits best given the
number of parameters and degrees of freedom. In the Bayesian approach this
amounts to the comparison of Bayesian evidence. For the least-squares parameter
fitting the equivalent is the comparison of the $\chi^2$ of the two models.

We therefore run LIL with and without the CO component. The difference in
$\chi^2$ between the two models again has a $\chi^2$ distribution with the
degree of freedom equal to one (i.e. difference in the degrees of freedom
between the two models) \citep[see e.g.][]{kendall}. We therefore accept the model with the CO component only if
this model gives an improvement of $\Delta \chi^2\ge 2.7$ over the model
without the CO component. This corresponds to a $10\%$ probability that we
will accept a model with the CO component when there is no CO component
present. When following this criteria the CO component is
now absent in $79.6\%$ of the pixels ($84\%$ if only considering pixels
outside the mask) which is a considerable improvement
over the single model approach. We will refer to this approach as LIL-MS
from now on.  There is  a small improvement in the $\chi^2$
distributions in LIL-MS compared to LIL (see section \ref{planckres} for a
plot of LIL-MS $\chi^2$ distributions).

The CMB residuals are shown in Fig. \ref{Fig:rescmbco} and the
corresponding normalized errors in Fig. \ref{Fig:stdco}. The Ecliptic
systematic features are invisible in the residual maps and are only visible
in the normalized error map because the error is underestimated in these
extremely low noise regions.  The systematic bias towards the positive
residuals around the galactic plane has also gone away and the residuals
everywhere except the highest foreground regions, which would be masked, are
consistent with Gaussian noise.

The residuals and standard deviation for the CO component is shown in
Fig. \ref{Fig:coresstd}. There is a marked improvement compared to LIL. Outside the
mask/Galaxy $84\%$ of the pixels have no detected CO component which is much closer
to reality. We show the PDFs for the foreground amplitudes for LIL-MS  in Fig. \ref{Fig:fgpdfco}. There is an improvement in
the CO component which has become closer to a Gaussian but not much change
in the other components. Note that the dust component is very lopsided
for the pixels inside the mask because the monopole that was subtracted in doing
this calculation was calculated from pixels at latitude
$|b|>30^{\circ}$. The biases for the dust component are different at high
and low latitudes since the low frequency component is very small at
$|b|>30^{\circ}$ where there is still considerable dust component. This
suggests that we should extend the model selection to the low frequency
component also. In particular if the model comparison favors absence of low
frequency component in part of the sky then we could remove the noisy
lowest two LFI
channels completely from the fit and fit a 4-parameter model without the LFI component
to the remaining five channels. We leave such an extension of our
model selection approach for future work. Since the systematic effect we
have removed is quite small, there is not much change in the PDFs for the
CMB component and we do not show them again.

\subsection{Resolution and noise in CMB maps}

The FFP6 simulations also provide half-ring maps in addition to the full
survey maps and we use these to estimate the resolution and noise in our
CMB maps. We have processed the half-ring frequency  maps with LIL-MS which produces
the corresponding two CMB half-ring maps. The half the difference between
the two CMB half-ring maps (HRHD) then gives an estimate of the noise in the
half ring half sum (HRHS) or the average of the two CMB half-ring maps. To
get an estimate of the resolution, we calculate the
pseudo-power spectrum ($\hat{C}_{\ell}$) of the HRHD and HRHS maps on masked
sky. We create the mask as described in section \ref{mask} but lower the
thresholds so that $30\%$ of the sky is masked and also apodize the mask
with a $\theta_{\rm ap}=30'$ Gaussian function in pixel space, replacing
the $1s$ in the mask by $1-\exp(-9\theta^2/(2\theta_{\rm ap}^2))$ for
$\theta\le\theta_{\rm ap}$, where $\theta$ is the distance of pixel from the
edge of the mask. We then deconvolve the mask by solving the
linear convolution equation for $C_{\ell}$ \citep[see][for a derivation]{hivon}
\begin{align}
\hat{C}_{\ell_1}=\sum_{\ell_2}M_{\ell_1\ell_2}C_{\ell_2}\label{Eq:dcon}
\end{align}
where $C_{\ell_2}$ is the full sky power spectrum and 
\begin{align}
M_{\ell_1\ell_2}=\frac{2\ell_2+1}{4\pi}\sum_{\ell_3}(2\ell_3+1)W_{\ell_3}\left(
\begin{array}{lcr}
\ell_1 & \ell_2 & \ell_3 \\
  0 & 0 & 0
\end{array}
\right)^2,
\end{align}
$W_{\ell}$ is the power spectrum of the mask and the term in brackets is
the Wigner-3j symbol. This equation is strictly applicable to only ensemble
averages but it is good enough for our purpose to test the quality of our
algorithm. Subtracting the HRHD or noise power spectrum $C_{\ell}^{\rm
  noise}$ from HRHS power spectrum $C_{\ell}^{\rm
 HRHS}$
gives an estimate of the CMB power spectrum which we call $C_{\ell}^{\rm
  out}$. We apply the same mask and deconvolution procedure to the input
CMB map. The input CMB map for the FFP6 simulation was made with a $4'$
Gaussian beam with spherical harmonic transform $b(l)$. We therefore divide
the power spectrum of the input CMB map by $b(l)^2$ to get the input CMB
power spectrum $C_{\ell}^{\rm in}$. The ratio of the output and the input power
spectrum then gives an estimate of the beam function of our CMB map and we
find it to be well approximated by a 

Gaussian beam of FWHM $7.8'$. This is close to
the resolution achieved by {\sc Commander-RULER} of $7.4'$
\citep{planckcomp}.  The noise in our CMB map is a little worse compared
to {\sc Commander-RULER}. The reason for it is probably the fact that we did not
change the resolution of all frequency maps to a common one. Therefore the
 high noise low resolution channels get higher weight than they are
 entitled to during the least squares fitting, i.e. the relative noise in
 the low resolution channels is underestimated compared to the higher
 resolution channels. This effect is aggravated for us compared to {\sc
   Commander-RULER} since we fit a non-linear model at full resolution while 
{\sc RULER} fits a linear model. {We make an  such an improvement in our method
 to mitigate the noise and improve the resolution in the next section.}
\begin{figure}
\resizebox{\hsize}{!}{\includegraphics{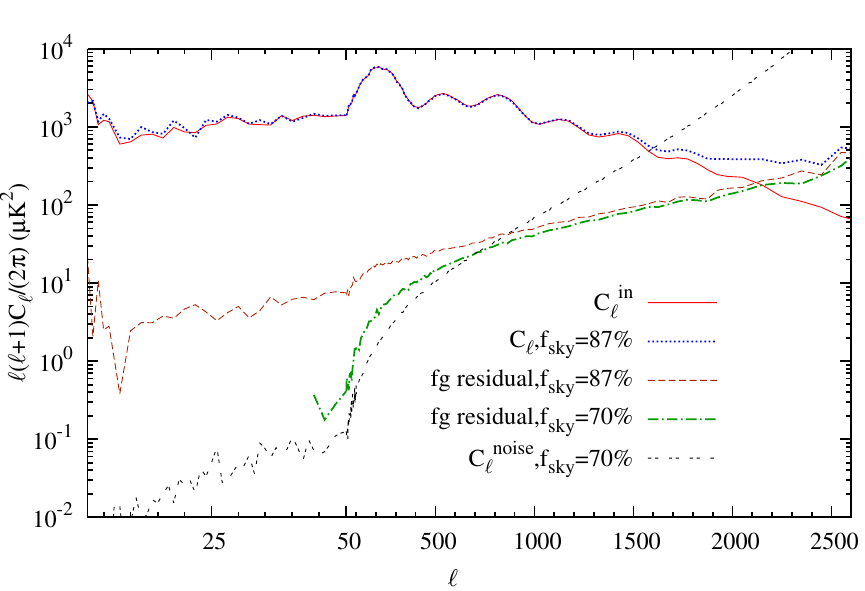}}
\caption{\label{Fig:cl70} The angular power spectra  for FFP6 simulations from the re-beamed LIL-MS
  algorithm corrected for the effect
  of mask (using $70\%$ and $87\$$ sky fractions) and beam. The power spectrum has been binned to make differences in different curves
  more visible. The noise calculated from the half-ring half-difference
  maps and the foreground residuals are also shown. All spectra except for the
  noise are the cross-spectra of half-ring maps and therefore without the
  uncorrelated white noise.}
\end{figure}

\subsection{Improvement in  resolution and noise properties by rebeaming}\label{sec:rebeam}
So far, we have worked with the maps at native resolution of different
Planck channels which vary from $\approx 32.2$ arcmin FWHM at 30 GHz to
$\approx 13.2$ arcmin
at 70 GHz for the LFI channels \citep{lfibeam}  and from $\approx 10$
arcmin at 100 GHz to $\approx 5$ arcmin at 353 GHz for the HFI channels
\citep{hfibeam}. This is not the optimal situation since we give the same
weight to the data in low resolution channels as the higher resolution
channels and the final resolution of the CMB map is close to the lowest
resolution among the channels with the least noise at native
resolution. The component separation methods which work in harmonic domain
are able to use information from the high resolution highest frequency
channels while down-weighting the low resolution channels on small
scales/high $\ell$. In particular they can change the weights given to
the information from different channels depending on the scales being
looked at and therefore achieve effective resolution corresponding to the
highest frequency channels, which are not completely dominated by foregrounds, used.

There is no reason however that we cannot apply the same principle when
working in pixel space. In particular to ensure that we assign correct
weights to information from different channels, we should process the maps
to have same resolution. The {\sc Commander} algorithm smooths all channels
to a resolution corresponding to the worst resolution channel (in practice a
little worse than that) which is the 30 GHz channel. We instead choose to
unsmooth or re-beam all HFI  channels to $5$ arcmin resolution corresponding to the
best HFI channels we are considering. This is similar to what is done in the
{\sc SMICA} and {\sc NILC} algorithms \citep{planckcomp}. The noise in the LFI
channels is too large for this rebeaming and we keep them at there native
resolution. This should not be a problem since because of the way LIL is
constructed, the information from these channels {is mostly  used to
constrain} the low frequency foreground component. We also add an additional
cut-off to the the spectrum
at $\ell=2800$ to avoid excessive small scale noise in the 100 GHz and 143
GHz channel. Rebeaming the 100 GHz and 143
GHz channels to 5 arcmin increases the noise per pixel and these channels
are therefore down-weighted (or correctly weighted with respect to the other
HFI channels), especially in the clean parts of the sky, when
doing the parameter fitting. We estimate the average white noise variance for the
re-beamed maps using the HRHD maps which together with the hit-count maps
gives us the variance maps to use in the LIL.
\begin{figure*}
\resizebox{\hsize}{!}{\includegraphics{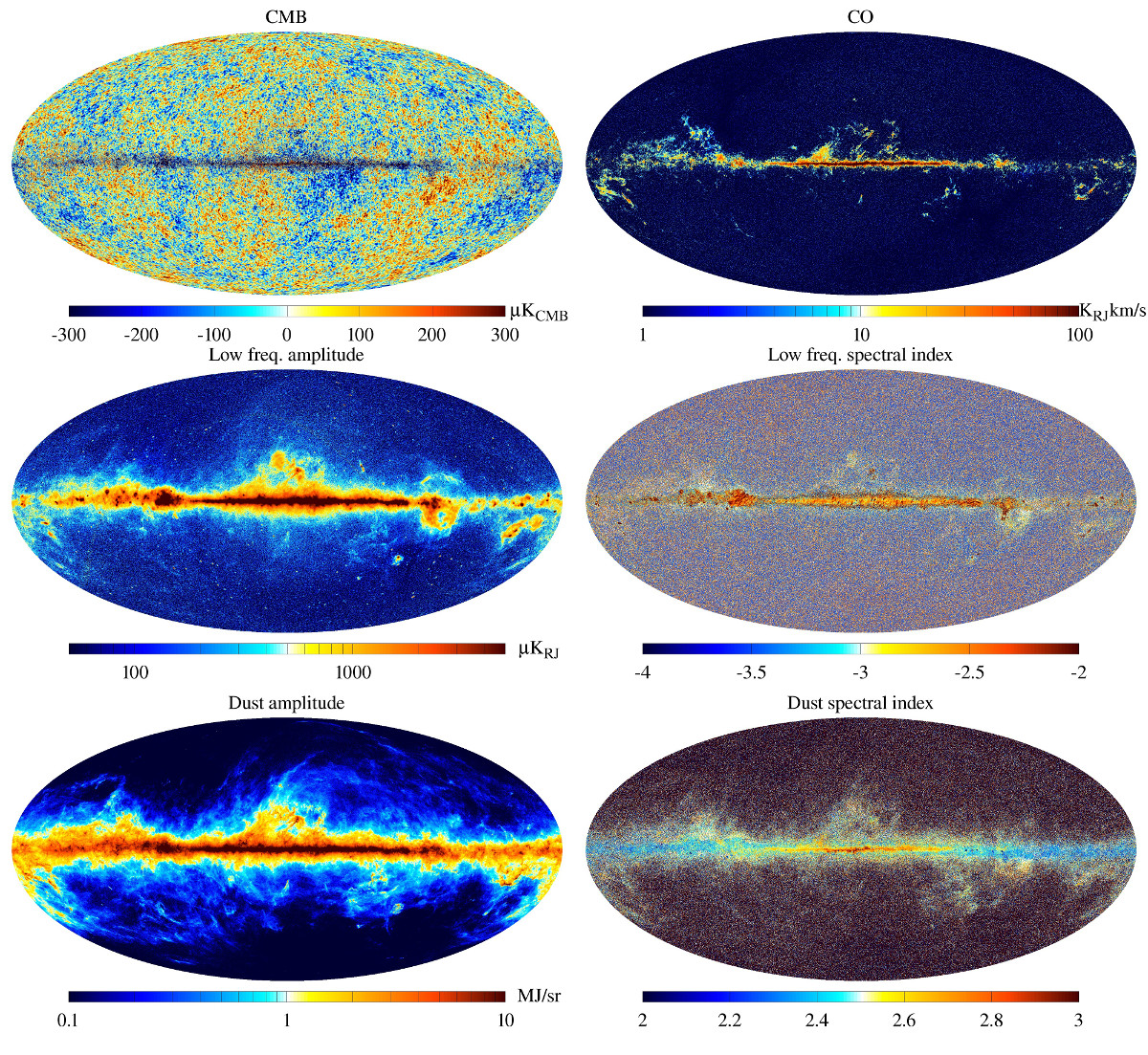}}
\caption{\label{Fig:planckcomp}The CMB and foreground parameter maps from
  LIL-MS applied to Planck data.}
\end{figure*}
\begin{figure*}
\resizebox{\hsize}{!}{\includegraphics{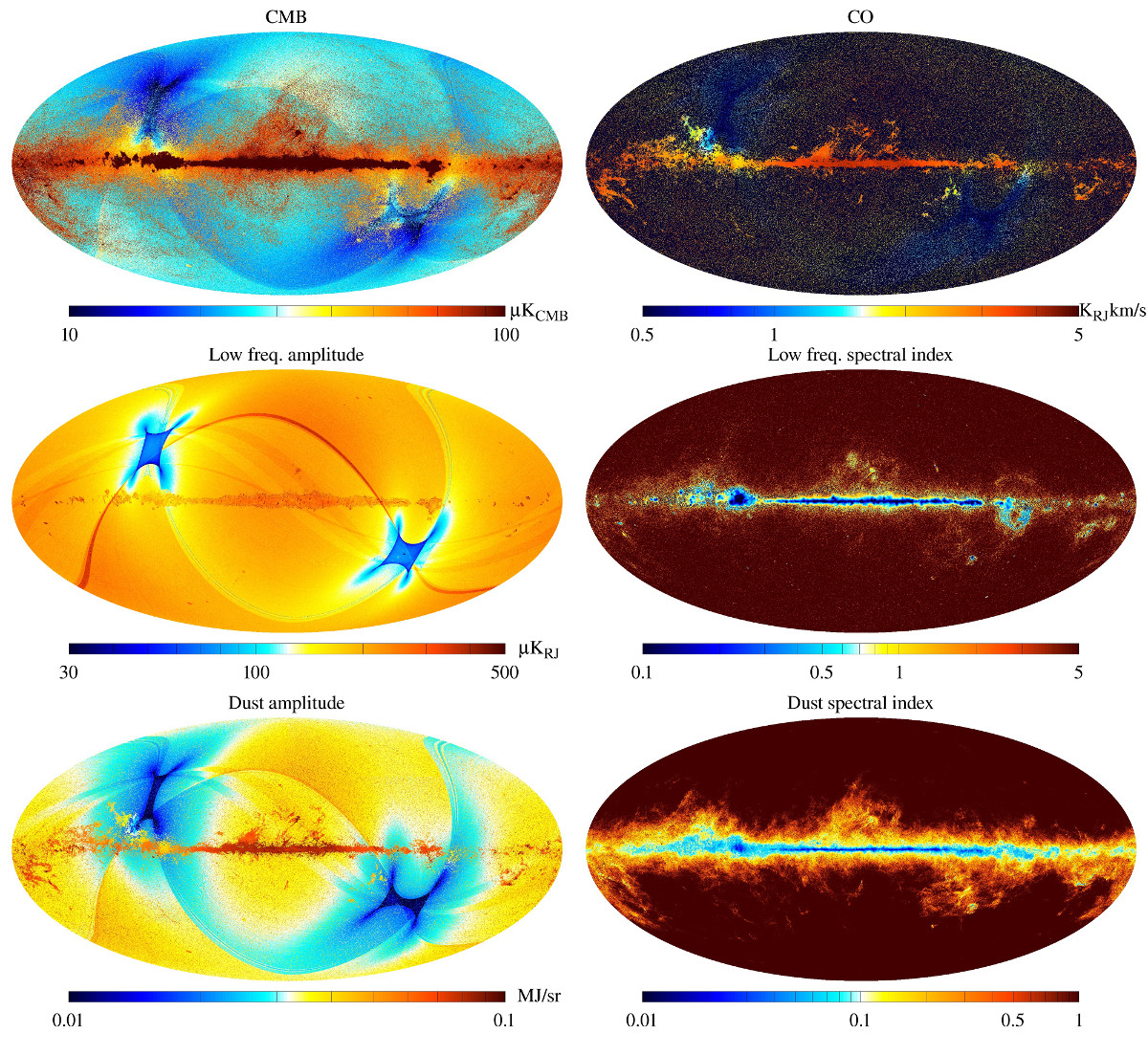}}
\caption{\label{Fig:planckcomperr}The CMB and foreground parameter standard
  deviation 
  maps estimated by 
  LIL-MS using Fisher matrix.}
\end{figure*}

{The effective beam of the resulting CMB
map from the re-beamed maps  is  $\approx 5.2$ arcmin.  The angular power
spectra for the HRHD, and the cross power spectra between the half-ring
maps for the CMB and the 
residual foregrounds  are shown in Fig. \ref{Fig:cl70} for
the $70\%$ mask and corrected with for the
effect of beam using the 5.2 arcmin FWHM Gaussian beam.  The $C_{\ell}$ have been averaged
in bins of size increasing in size with $\ell$  to make the curves smoother and
easier to interpret.  The residual
foregrounds are below the signal at $\ell\lesssim 2000$ and negligible on
large scales. The shape and amplitude of the residuals is similar to those
obtained by \citet{planckcomp} for similar sky coverage. We also show in
Fig. \ref{Fig:cl70} the same power spectra but for a smaller mask, masking
only $\approx 13\%$ of the sky. Even with this considerable larger sky
fraction, the foregrounds are still sub-dominant at $\ell<2000$ and there
is only a small increase in the  foreground residuals. We have used an
expanded linear scale at $\ell<50$ to show the small $\ell$ features. At
small $\ell$ the features are dominated by the deconvolution errors, as our
simple deconvolution algorithm \citep{hivon} is not expected be accurate on
large angular scales. This is obvious in the $70\%$ residual spectrum where
the deconvolved power spectrum solution to Eq. \ref{Eq:dcon} becomes negative at $\ell<40$.
}

\begin{figure}
\resizebox{\hsize}{!}{\includegraphics{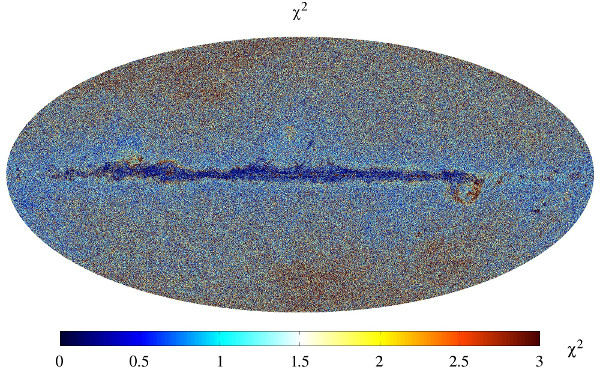}}
\caption{\label{Fig:chisqplanck} The  $\chi^2$ map for the least squares fit performed
  by LIL-MS. For 1 or 2 degrees of freedom that we have with or without the CO
component, the average $\chi^2$ is expected to be between 1 and 2.}
\end{figure}

\section{Application to Planck data and comparison with official Planck
  maps}\label{planckres}
{We now apply LIL-MS to the Planck nominal and full data releases.}
The maps for the nominal mission  for 6
parameters recovered by LIL-MS are shown in Fig. \ref{Fig:planckcomp}. The
CO amplitude is negligible away from the galactic plane and model selection
plays an important role in recovering the true behavior of CO. In
particular the CO component was removed from $84\%$ of the pixels outside a
$12.3\%$ mask on the highest foreground regions. The low frequency component
parameters are consistent with those from {\sc Commander}. The low
\begin{figure*}
\resizebox{\hsize}{!}{\includegraphics{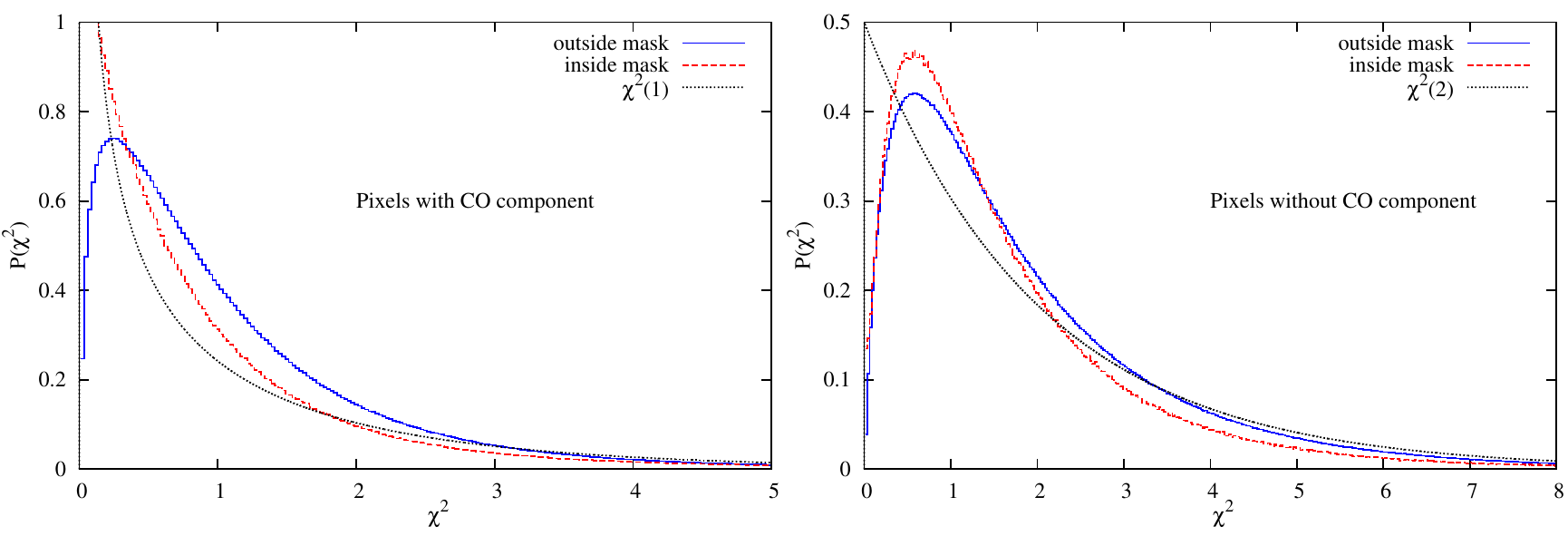}}
\caption{\label{Fig:planckchi} The $\chi^2$ probability density function
  obtained by LIL-MS for Planck maps and compared with the theoretically expected $\chi^2$
  distributions with 1 and 2 degrees of freedom. The distributions are
  close to what we predicted from the FFP6 simulations. The tails for real data has even better
  agreement with the theoretical distributions than what we had for simulations.}
\end{figure*}
frequency amplitude is mostly determined by 30 GHz channel which has high
signal to noise and this is apparent in the amplitude map. To constrain the
spectral index we need high signal to noise in more than one channel. Since
the foreground amplitude drops quite a bit from 30 GHz to 44 GHz, the
spectral index is well constrained in much smaller region and therefore
looks much noisier at high galactic latitudes. The dust amplitude is very
well constrained and fine features such as streams and filaments are well
recovered over most of the sky. Same constraints as low frequency index
apply also to the dust spectral index and it is well constrained in regions
of high amplitude but becomes noisier and unconstrained as we go to the
higher latitudes. Our CMB map has some residual small scale noise in a
narrow ridge in the
galactic plane compared to the CMB maps released by the Planck
collaboration. This is probably because we did not adjust the resolution of
channel maps to a common resolution before parameter fitting. We will discuss
a possible solution to this in the conclusions section.

We show in Fig. \ref{Fig:planckcomperr} the standard deviation estimated
using the Fisher matrix. Comparison of these maps gives useful information
about the degeneracies between the parameters. The CMB is influenced by both
the dust and the low frequency components and against a background that
follows the Planck hit count map, the influence of low and high frequency
foregrounds is clearly visible. Comparison of CO map with dust map shows
that these two are very degenerate while there is not much degeneracy
between the low frequency foregrounds with the dust and CO. The spectral
index is well constrained in the high signal regions, where there is
good signal to noise for the corresponding foreground component in at least two of the
channels. The errors on the spectral indices therefore follow the morphology
of the respective foreground components.

We show the $\chi^2$ map in Fig. \ref{Fig:chisqplanck} and compare the
$\chi^2$ distribution with expected distributions with one or two degrees
of freedom in Fig. \ref{Fig:planckchi}. The tails agree very well with the $\chi^2$ distribution for one degree of freedom when the CO component is included and with the $\chi^2$ distribution for two degrees of freedom when the CO component is excluded by model selection. Overall the distributions are close to what we expected
from simulations.
\begin{figure*}
\resizebox{\hsize}{!}{\includegraphics{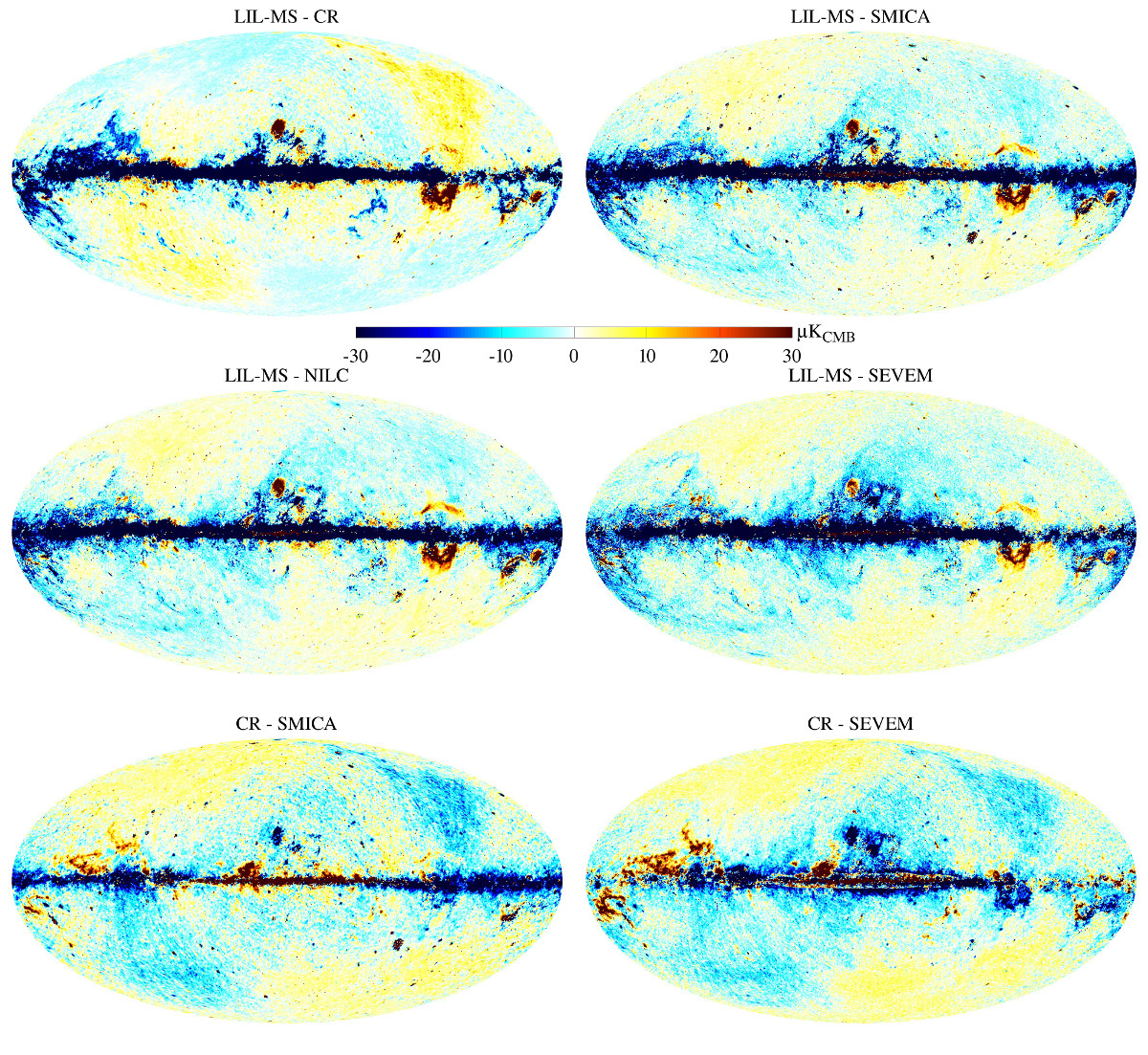}}
\caption{\label{Fig:lil-planck} Difference between LIL-MS CMB map and maps
  released by the Planck collaboration using different algorithms. All maps
  have been downgraded to nside=128. We also show for comparison the
  difference between the {\sc Commander-RULER} and {\sc SMICA}/{\sc SEVEM}
  maps on the same color scale. A monopole and
  dipole calculated at latitudes $|b|>30^{\circ}$ has been subtracted.
}
\end{figure*}

\subsection{Comparison with the Planck collaboration results}
\begin{figure}
\resizebox{\hsize}{!}{\includegraphics{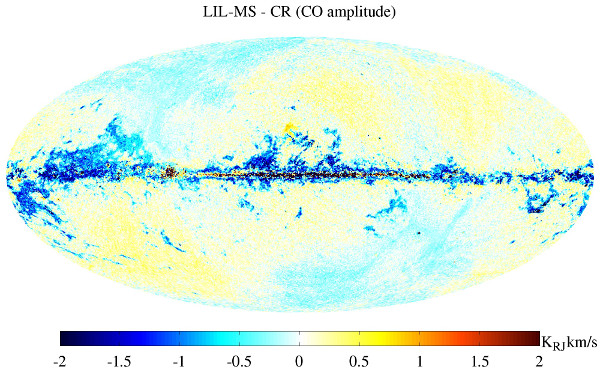}}
\caption{\label{Fig:lil-CRco} Difference between LIL-MS CO amplitude at
  100 GHz  and the {\sc CR} map degraded to nside=128. A monopole and
  dipole calculated at latitudes $|b|>30^{\circ}$ has been subtracted.
}
\end{figure}

\begin{figure*}
\resizebox{\hsize}{!}{\includegraphics{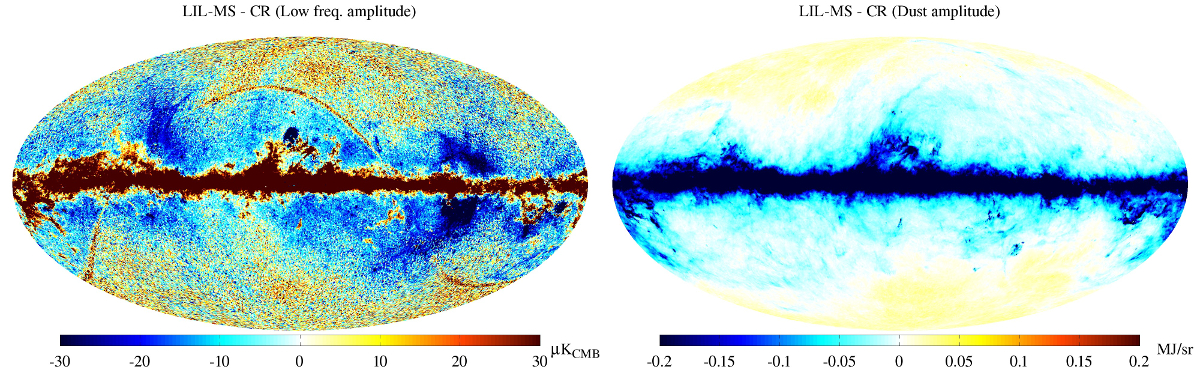}}
\caption{\label{Fig:lil-CRfg} Difference between LIL-MS low frequency and
  dust amplitudes at
  30  GHz  and 353 GHz respectively and the corresponding {\sc CR} maps degraded to nside=128. A monopole and
  dipole calculated at latitudes $|b|>30^{\circ}$ has been subtracted.
}
\end{figure*}

We show in Fig. \ref{Fig:lil-planck} the difference between our CMB maps and the CMB maps
from different algorithms released by the Planck collaboration
\citep{planckcomp}. We also show for comparison the difference between the
{\sc Commander-RULER}({\sc CR}) and {\sc SMICA}/{\sc SEVEM} maps on the
same color scale. The agreement between our maps and other algorithms is
as good as the agreement in-between the methods used by the Planck
collaboration. All maps are downgraded to HEALPix nside=128 so they can be
compared with similar maps in \citet{planckcomp}. We also show the comparison between the foreground
amplitudes recovered by us and those from the {\sc CR} in
Fig. \ref{Fig:lil-CRco} for CO and Fig. \ref{Fig:lil-CRfg} for the low
frequency and dust amplitudes. The agreement is
again quite good away from the galactic plane. It is not possible to directly compare the spectral
indices since for {\sc Commander} algorithm they are calculated at much lower
resolution and are mostly driven by the prior over most of the sky away
from the Galactic plane. In the Galactic plane we see that our maps are
broadly consistent with those from {\sc Commander} by comparing our
Fig. \ref{Fig:planckcomp} with the corresponding figure in
\citet{planckcomp}. {The difference away from the galactic plane follow
dust emission and may arise because we turn off the CO component when our
algorithm decides that it is absent while {\sc CR} fits for it
everywhere. Also {\sc Commander} has much tighter priors on spectral
indices compared to the constraints used by LIL.}

\begin{figure}
\resizebox{\hsize}{!}{\includegraphics{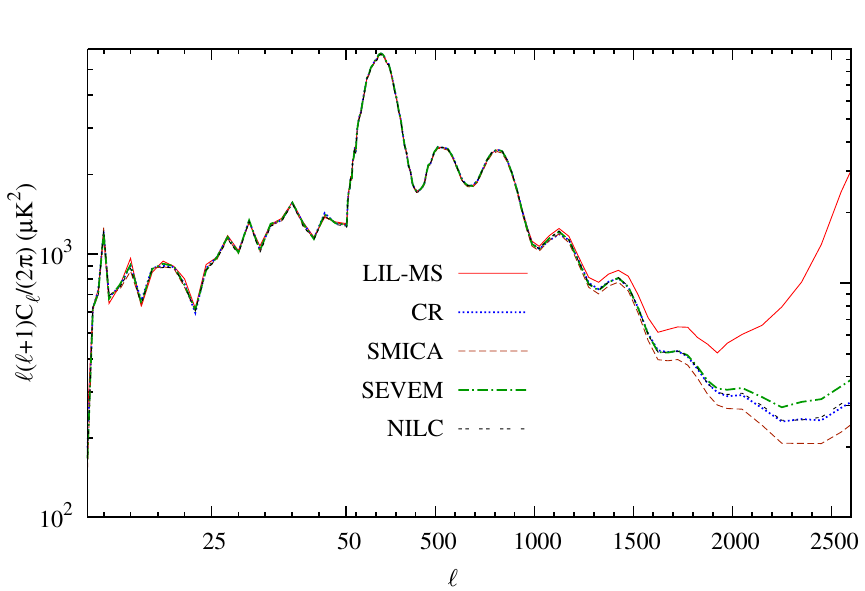}}
\caption{\label{Fig:clsky} Comparison of power spectrum from LIL with the maps
  released by the \citet{planckcmb2015} for full mission data release for
  the same $87\%$ mask. All spectra are the cross-spectra of the half-ring
  maps calculated with our code and corrected for the effect of mask
  \citep{hivon} and
  beam ($5'$ FWHM for Planck released maps and $5.2'$ for LIL). An expanded
linear scale is used at low multipoles.}
\end{figure}

{In Fig. \ref{Fig:clsky} we compare the CMB power spectra from LIL-MS with
the CMB maps released by Planck collaboration for the full mission
release \citep{planckcmb2015}. We present the cross-spectra between half-ring maps so that we do
not have to subtract the noise separately. The same sky fraction of $87\%$
is used for all curves and the effect of masks and beam have been deconvolved. The power
spectra are especially useful for comparing small scales where we see that
LIL-MS CMB maps have higher noise compared to the other methods. The excess
power on small scales is similar to what was present in the Commander-Ruler
maps in the nominal mission maps \citep{planckcomp} and is mostly coming
from the noisy low frequency channels. In the full mission
data release Planck collaboration uses an approach of fitting the
parametric model at many different resolutions, using all channels on
low resolutions and only the less noisy HFI channels at higher
resolutions \citep{planckcr2015}. This multi-resolution approach allows
them to decrease their noise on small scales to the levels seen by the
other algorithms. The same approach can in principle be followed also for
our algorithm, we however defer this and other improvements to a future publication.}

\section{Conclusions}
The main aim of this paper is to  present a least squares parameter fitting
algorithm, LIL,  optimized for component separation in the CMB sky. Our
algorithm is extremely efficient, fitting 6 parameter model to 7 frequency
channels for 50 million pixels in 160 CPU-minutes or a few minutes running
in parallel on few tens of cores. We have
also argued for an extension to the algorithm, LIL-MS, by including model
selection for the components such as CO and perhaps also for the low frequency component, which
we know from observations are only present in detectable amount over a
fraction of the sky. We have shown that non-linear parameter fitting can
be done at high resolution. The main aim of developing such a parameter fitting algorithm is
to try to get maximum information about the foregrounds and the CMB from
data while making least amount of assumptions. In particular our
assumptions and models are motivated by  the prior knowledge about the foregrounds from
other observations. Our method is still not optimal and there are several
improvements which can be done based on the results we have obtained so
far.

\begin{enumerate}
\item Difference in resolution between different channels can be a source
  of high frequency noise and we see this in the CMB map. In addition the
  resolution of our final CMB map is close to that of 143 GHz
  channel. {\sc SMICA} and {\sc NILC} \citep{planckcomp} achieve a higher
  resolution by re-beaming all channel maps to $5'$. In principle we see no
  reason why we cannot do that also for LIL-MS. Re-beaming to highest
  resolution channel will lead to increase in noise in the lower resolution
  channels. Thus during parameter fitting, the low resolution data will get
  the correct 
  additional down-weighting and the final maps would be closer to the
  resolution of the highest resolution channels, especially for the
  CMB. We showed that this is indeed the case in section
  \ref{sec:rebeam}. {A multi-resolution approach similar to \citet{planckcr2015}
  should yield further improvements on small scales.}
\item The low frequency component, like the CO component, is important only
  on a fraction of the sky. This suggests to extend the model selection to
  include the low frequency component. In particular in parts of the sky
  where  a model with a low frequency component is disfavored, the two lowest
  frequency LFI channels can simply be omitted from the fit.
 \item If we are interested in low frequency component, then doing a fit at
   the nside=2048 does not really make sense. For the low frequency
   component therefore a dedicated analysis can be done by re-beaming to  a
   lower resolution close to that of the 70 GHz channel.
\item We have assumed that the noise between different pixels is
  uncorrelated at nside=2048. This is not true since for most channels the
  beam size is much bigger than the pixel size. A better treatment of noise
  is therefore needed when doing cosmological analysis. We note that {\sc
    Commander} also ignores correlations between the pixels.\
\item {Our algorithm, since it gives a quantitative estimate of goodness of
  fit, is most useful for the rare and weak signals such CO line emission and SZ
  effect. The application to the construction of full sky map of SZ effect
  is presented in separate publications \citep{k2015,ks2015} with emphasis
  on separating the CO emission contamination from the SZ signal.}
\end{enumerate}

We have shown that our simple approach to component separation using the least squares fitting works quite well
for multi-frequency experiments like Planck. An advantage of
parameter fitting over other methods \citep[see however][]{bayessmica} is that the errors we get take into
account the uncertainties in the foregrounds and that it allows for
considerable flexibility to the foregrounds to vary over the sky. 
We have shown our results to be  consistent with
the published results  by the Planck collaboration. We hope to overcome the
shortcomings in our algorithm concerning the resolution and noise compared
to the existing methods in the near future. Our algorithm can be
 extended to include polarization, since the only requirement for it
to be applicable is that the foreground model be almost-linear. As recent
results from the Bicep2 show \citep{bicep2,planckbicep} we are entering a regime in the CMB
experiments where the signal is buried in the foregrounds and assumptions
about the foregrounds can have a big influence on the interpretation of the
experimental data. 
\section*{Acknowledgements}
This paper used observations obtained with Planck
(\url{http://www.esa.int/Planck}), an ESA science mission with instruments and
contributions directly funded by ESA Member States, NASA, and Canada. We
 also acknowledge  use of the HEALPix software
 (\url{http://healpix.sourceforge.net}) and FFP6 simulations generated
 using  the Planck
 sky model
 \url{http://wiki.cosmos.esa.int/planckpla/index.php/Simulation_data}. I
 also thank Eugene Churazov for useful  discussions on the component separation methods.

\bibliographystyle{mn2e}

\bibliography{lilcmb}

\appendix

\section{Total foreground residuals in all the seven Planck channels used in component
  separation}

\begin{figure}
\resizebox{\hsize}{!}{\includegraphics{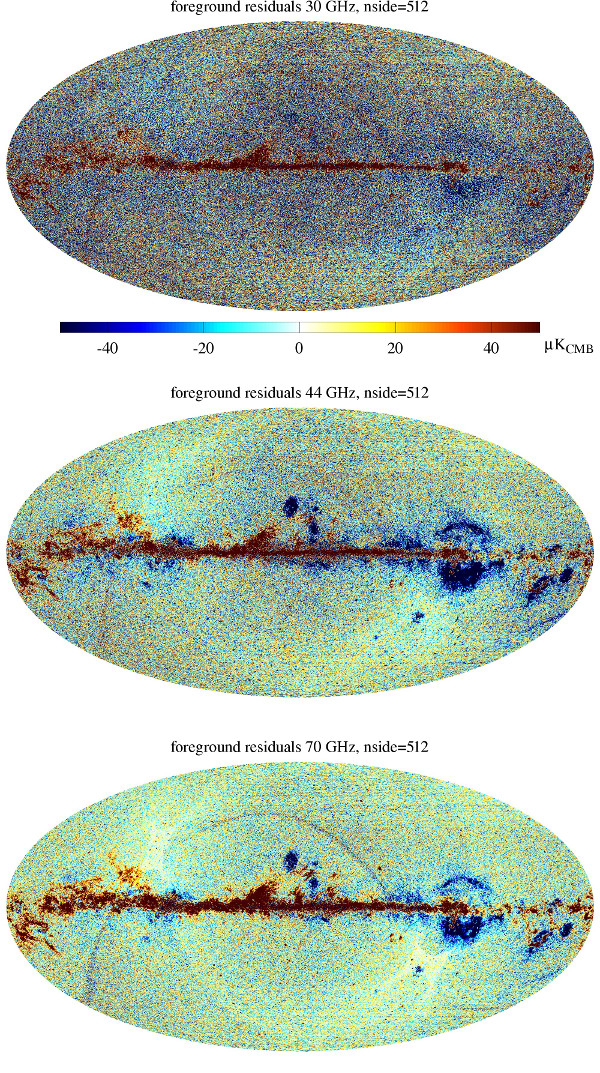}}
\caption{\label{Fig:fgreslfi} Residuals for the sum of all foregrounds for
  the LFI channels  in FFP6 simulations.
}
\end{figure}

\begin{figure*}
\resizebox{\hsize}{!}{\includegraphics{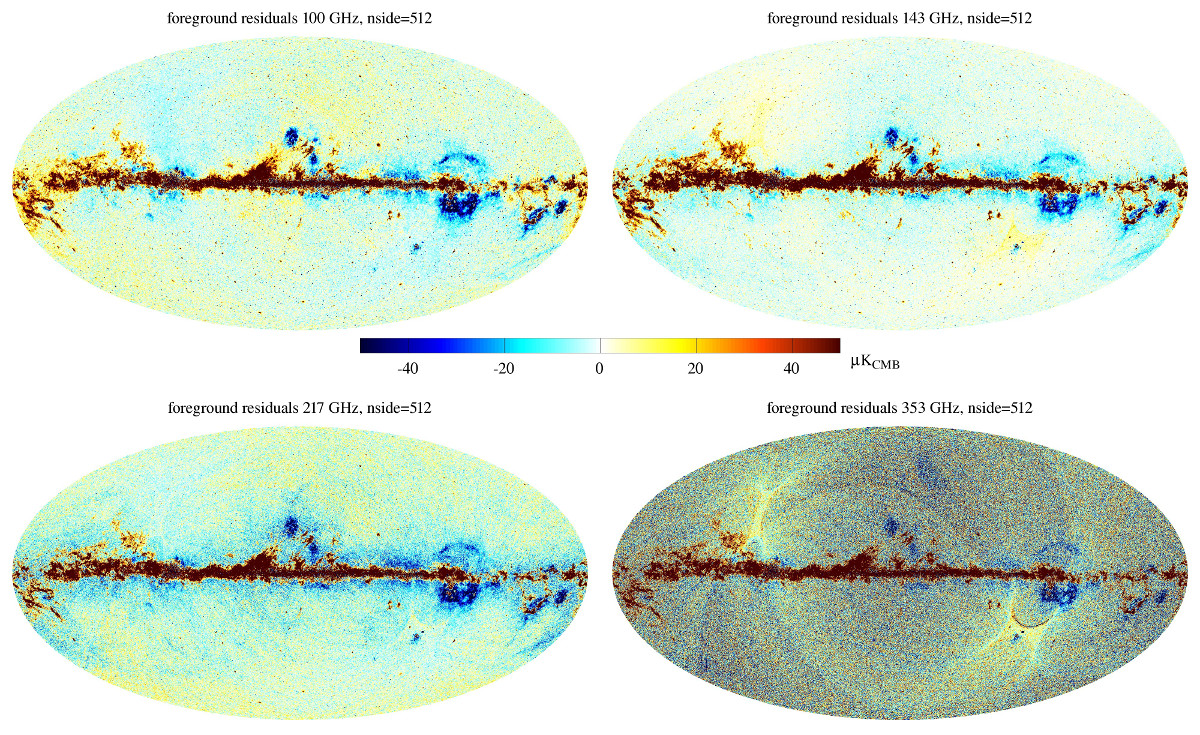}}
\caption{\label{Fig:fgreshfi} Residuals for the sum of all foregrounds for
  the HFI channels in FFP6 simulations.
}
\end{figure*}

We show in Figs. \ref{Fig:fgreslfi} and \ref{Fig:fgreshfi} the residuals
for the sum of all foreground components in order to validate the recovered
values of the spectral indices.

\pagebreak

\end{document}